\documentclass[twocolumn,floatfix,superscriptaddress,aps,prb]{revtex4-2}

\usepackage[utf8]{inputenc}
\usepackage{graphicx}
\usepackage{amsmath,mathtools}
\usepackage{amssymb}
\usepackage{bm}
\usepackage{hyperref}
\usepackage{color}
\usepackage{bbold}
\usepackage{lineno}
\usepackage[normalem]{ulem}

\DeclareMathOperator{\sgn}{sgn}

\DeclareMathOperator{\Imm}{Im}

\newcommand{\addYW}[1]{\textcolor{blue}{#1}}

\usepackage{color}
\usepackage{amssymb}

\newcommand{\beq} {\begin{equation}}
\newcommand{\eeq} {\end{equation}}
\newcommand{\bea} {\begin{eqnarray}}
\newcommand{\eea} {\end{eqnarray}}
\newcommand{\be} {\begin{equation}}
\newcommand{\ee} {\end{equation}}
\renewcommand{\(}{\left(}
\renewcommand{\)}{\right)}
\renewcommand{\[}{\left[}
\renewcommand{\]}{\right]}

\def \beq {\begin{equation}}
\def \eeq {\end{equation}}
\def \beqa {\begin{eqnarray}}
\def \eeqa {\end{eqnarray}}
\def \bseq {\begin{subequations}}
\def \eseq {\end{subequations}}

\begin{document}

\title{Solvable model for a charge-$4e$ superconductor}

\author{Nikolay V. Gnezdilov}
\affiliation{Department of Physics, University of Florida, Gainesville, Florida 32601, USA}

\author{Yuxuan Wang}
\email{yuxuan.wang@ufl.edu}
\affiliation{Department of Physics, University of Florida, Gainesville, Florida 32601, USA}

\begin{abstract}
A charge-$4e$ superconductor forms due to the condensation of quartets of electrons. While in previous works the mechanism for the formation of charge-$4e$ superconductivity has been analyzed in terms of the binding of Cooper pairs in unconventional superconductors, its properties in the fermionic sector have not been studied systematically due to its inherently interacting nature even at the mean-field level. Here we propose a solvable model for a charge-$4e$ superconductor -- a spinful version of the Sachdev-Ye-Kitaev model with an  anomalous quartic  term. We show that the ground state is gapless and resembles a heavy Fermi liquid.
We analytically solve for the superfluid density and show that it is perturbative in the strength of the charge-$4e$ order parameter, in sharp contrast with a regular (charge-$2e$) superconductor. Upon lowering temperature, we show that the correlation between charge-$4e$ order and regular interaction terms can drive a first-order phase transition to a charge-$2e$ superconducting state.
\end{abstract}

\date{\today}

\maketitle




\section{Introduction}
\label{sec:intro}
As described by the Bardeen-Cooper-Schrieffer (BCS) theory, a metallic system becomes superconducting when Cooper pairs with charge $2e$ formed by electrons condense~\cite{Bardeen1957Theory}. Going beyond this paradigm, it has been recently theoretically proposed that in the presence of strong correlation effects, a condensate of a quartet of electrons with charge $4e$ can form and condense. Such a state is known as a charge-$4e$ superconductor, which has been theoretically proposed to exist in a range of strongly correlated systems~\cite{Kivelson1990Doped,  Wu2005Competing, Berg2009Charge, Berg2009Theory, Radzihovsky2009Quantum, Herland2010Phase, Moon2012Skyrmions, Jiang2017Charge, Agterberg2020Physics, Fernandes2021Charge4e, Jian2021Charge}. {The charge-$4e$ condensate 
was also proposed to exist in superfluid $^3$He~\cite{Volovik1992Exotic}.} 

Despite its resemblance to a regular charge-$2e$ superconductor, the theoretical description of a charge-$4e$ superconductor is much more challenging. First, unlike charge-$2e$ superconductivity that emerges for weak coupling via a logrithmically divergent contribution to pairing susceptibility~\cite{Abrikosov1975Methods}, no such weak-coupling instability of electrons exists towards charge-$4e$ superconductivity. For this reason, existing theories of charge-$4e$ superconductivity usually assume some underlying strong interactions between the electrons, and address the formation of charge-$4e$ superconductivity within a bosonic theory describing the binding of two Cooper pairs and the condensing of the composite object~\cite{Berg2009Charge, Berg2009Theory, Herland2010Phase, Fernandes2021Charge4e, Jian2021Charge}. (See
also Ref.~\cite{Babaev2004Superconductor} for an example in a neutral superfluid from a composite order parameter.)
As an emergent degrees of freedom, the Cooper pair order parameter breaks particle-number conservation as well as some other symmetries such as translational symmetry (known as a pair-density-wave state)~\cite{Berg2009Theory,Agterberg2020Physics}, rotational symmetry~\cite{Fernandes2021Charge4e,Jian2021Charge},  or certain internal symmetries~\cite{Herland2010Phase}.  
Upon increasing temperature, charge-$4e$ condensate emerges as a vestigial order~\cite{Babaev2004Superconductor,Berg2009Theory,Fernandes-2012,Nie2014Quenched,Wang-2014} via partial melting of the Cooper pair order parameter by restoring certain spatial or internal symmetries. {

However, the behavior in the fermionic sector in a charge-$4e$ superconductor, which is  outside the scope of these approaches, remains to be understood. Unlike a charge-$2e$ superconductor which admits a non-interacting mean-field description, the mean-field description of a charge-$4e$ superconductor is inherently interacting, since the order parameter couples to four fermion operators~\cite{Jiang2017Charge}. While a charge-$2e$ order parameter gaps the Fermi surface, it is natural to expect that the charge-$4e$ order parameter leaves the Fermi surface gapless~\cite{Berg2009Charge} just like a regular interaction term in the Fermi liquid theory (although such an interaction breaks $U(1)$ symmetry).

Moreover, it is an open question how ``good" a superconductor a charge-$4e$ superconducting state is, i.e., whether it has a significant superfluid density, which is relevant for the Meissner effect and for its stability against phase fluctuations (especially in 2d). It is well-known that for a charge-$2e$ superconductor, due to the  gapped Fermi surface, the superfluid density is equal to the total electron density at low temperatures, independent of the magnitude of the order parameter (see e.g., Ref.~\cite{Altland2010Condensed} and Appendix \ref{app:ward_BCS}). As the fermions are expected to be gapless in a charge-$4e$ superconductor, a natural question is whether the superfluid density is comparable to the total electron density.

Upon lowering the temperature, the gapless excitations in the fermionic sector strongly contribute to the renormalization of the boson theory, which can potentially lead to the low-temperature instabilities such as the formation of charge-$2e$ condensate, consistent with results from the bosonic perspective. Indeed, in a determinant quantum Monte Carlo study, it was shown that the gapless fermionic excitations in general destroys the charge-$4e$ condensate in the presence of an attractive interaction at low temperatures~\cite{Jiang2017Charge}. However, an analytical understanding of such a pairing transition and the stability of the charge-$4e$ superconductivity is still missing.

The study of properties of a charge-$4e$ superconductor is also motivated by the theoretical description of a $\mathbb{Z}_4$ spin liquid with an emergent spinon Fermi surface. In terms of the fermionic spinons $f$, the system lacks the $U(1)$ symmetry but instead has a $\mathbb{Z}_4$ symmetry, which allows an  $\sim f^\dagger f^\dagger f^\dagger f^\dagger$ term, which is marginal under renormalization group flow. resembling a charge-$4e$ superconductor. As this interaction is marginal, it has been speculated~\cite{Barkeshli2013Gapless} that such a $\mathbb{Z}_4$ spin liquid admits a gapless Fermi-liquid like ground state, making it a promising candidate for gapless spin liquids. From this perspective, a detailed description of a charge-$4e$ superconductor, in particular whether its ground state is indeed gapless and Fermi-liquid like, is highly desirable.  

As we mentioned, an obvious challenge to the theoretical description of a charge-$4e$ superconductivity in the fermionic sector is its inherently interacting nature, even at the mean-field level. If the charge-$4e$ (``quartetting") order parameter is weak, one may expect a perturbation theory departing from free electrons to apply. However, as charge-$4e$ superconductivity is expected to derive from strong correlation effects, there is no particular reason to expect the charge-$4e$ order parameter to be weak compared to other  energy scales in the problem.  To this end, we note that there has been remarkable progress achieved by applying the Sachdev-Ye-Kitaev (SYK) model~\cite{Sachdev1993Gapless,Kitaev2015Simple} to 
describe strongly interacting fermionic systems without long-living quasiparticle excitations~\cite{Patel2019Theory,Hartnoll2018Holographic}.  The SYK model is a solvable of strongly interacting random fermions without quasiparticles in zero spatial dimensions~\cite{Maldacena2016Remarks,Kitaev2018Soft}. The generalizations of the SYK model to nonzero dimension lattice models predict the linear in temperature resistivity~\cite{Son2017Strongly,Chowdhury2018Translationally,Patel2019Theory}, which is the characteristic property of a strange metal~\cite{Takagi1992Systematic,Taillefer2010Scattering,Keimer2015Quantum}. On the other hand, SYK-like models have been constructed to analyze the superconducting transition of non-Fermi liquids~\cite{Patel2018Coherent,Esterlis2019Cooper,Cheipesh2019Reentrant,Chowdhury2020Intrinsic,Wang2020Solvable,Hauck2020Eliashberg,Hurtubise2020Superconducting,Phillips2020Exact,Wang2021Phase,Salvati2021Superconducting}. 

In this work, we present a solvable strong-coupling model for fermion-sector properties of a charge-$4e$ superconductor. Similar to the SYK model, the model is dominated by interaction effects, which in this case is due to the  condensate of the charge-$4e$ superconducting order parameter. Instead of speculating on the microscopic origin of the charge-$4e$ condensate, we treat it as a mean field and analyze the properties of the ground state and its low-temperature stability. Specifically, we consider a two-dimensional  itinerant-fermion system subject to local four-fermion interactions that are random in the flavor space but preserves spatial translation. As with all SYK-like models~\cite{Son2017Strongly,Chowdhury2018Translationally,Chowdhury2020Intrinsic}, we take the limit in which there are a large number of fermion flavors $N\to\infty$. In our model there are two types of interaction terms --- a ``regular" four-fermion interaction $\sim \mathcal{J}$ that descibes a pair-hopping process in flavor space within a lattice site, and an anomalous charge-$4e$ interaction $\sim \Delta_{4e}$ that describes a ``pair-pairing" process. The ratio of the strengths of the two interaction terms can be tuned, and we mainly focus on the nontrivial limit in which these interactions are much larger than the Fermi energy $\varepsilon_F$.

By solving the Schwinger-Dyson equations, we first show that the ground state of the charge-$4e$ superconductor is gapless. In fact, despite a non-conserving particle number, for this particular model there exist a Luttinger-Ward functional~\cite{Luttinger1960Ground} and the Fermi surface encloses a fixed volume equal to the expectation value of the number density. 
At lowest temperatures, the system behaves like a heavy Fermi liquid, in sharp contrast with a charge-$2e$ superconductor. Just like a Fermi liquid, the long-lived gapless quasiparticle is due to kinematic constraints of scattering processes in the vicinity of the Fermi surface. Nevertheless, we show that the system has a nonzero superfluid density given by $n_s/n=\beta\Delta_{4e}^2/[\mathcal{J}^2+(\beta+1)\Delta_{4e}^2]$. Here $\beta\equiv 4 \partial_k \Sigma(k_F)/v_F$ is a parameter responsible for the Fermi velocity renormalization, where $v_F$ is the Fermi velocity and $\partial_k \Sigma(k_F)$ is the momentum derivative of the self-energy at the Fermi level. Within our model we find $0<\beta=\mathcal{O}(1)$ and its numerical value depends on the details of the Fermi surface. 
In the  $\Delta_{4e}\ll \mathcal{J}$ limit, similar to the ``weak-pairing" limit for a charge-$2e$ superconductor, the superfluid density is vanishingly small. In the opposite limit, the superfluid density is a significant portion of the total electron density, but is still less than the latter. Similar to the lattice generalization of the SYK model~\cite{Son2017Strongly,Chowdhury2018Translationally}, the heavy-Fermi liquid behavior specifies an intermediate energy scale, which corresponds to the renormalized Fermi energy ${ \varepsilon_F^*}\propto \varepsilon_F^2/\sqrt{\mathcal{J}^2+\Delta_{4e}^2}$. At temperatures ${\varepsilon_F^*} \ll T\ll (\mathcal{J, K})$, the system behaves as a non-Fermi liquid. In this regime, the Fermi velocity is suppressed, i.e., $\beta=\mathcal{O}(\varepsilon_F/U)$. As a consequence, the superfluid density is parametrically small, $n_s/n=O(\varepsilon_F/U)$. 

{We argue that many results above traces back to the breaking of $U(1)$ symmetry in the mean-field ground state. This naturally raises the question of total charge conservation and, therefore, its relevance to an  isolated superconductor with fixed number of particles~\cite{Leggett1991Concept}. We show that particle-number conservation can be restored by treating the order parameter, in particular its phase degree of freedom, as a dynamical field. The ground state for such a system has strong entanglement in the Fock spaces of the fermions and bosons. We show that this makes our results, such as the violation of the Luttinger's theorem and superfluid density, valid for a $U(1)$-symmetric charge-$4e$ superconductor with fixed particle number.}

We also study the low-temperature stablity of the charge-$4e$ superconducting phase. This instability is akin to the pairing instability of a regular metal, which lowers its entropy by gapping out the Fermi surface upon lowering the temperature. For our model, we show that, through a first-order transition, the $4e$ bound state is unstable toward dissociation into equal-flavor, spin-singlet, and local Cooper pairs if the pair-hopping interaction $\mathcal{J}$ and the charge-$4e$ interaction $\Delta_{4e}$ are correlated. 
The transition temperature is determined by the strength of the correlation between $\mathcal{J}$ and $\Delta_{4e}$. Depending on whether the transition occurs in the heavy Fermi liquid regime or the non-Fermi liquid, the effective pairing interaction is either a constant analogous to the BCS theory, or logarithmically singular analogous to that in color superconductivity~\cite{Son1999Superconductivity,Chubukov2005Superconductivity}, described by the $\gamma$-model at $\gamma\to0_+$~\cite{Chubukov2021Pairing}. 
On the other hand, if there are no correlations between pair-hopping $\mathcal{J}$ and pair-pairing $\Delta_{4e}$, the system remains stable and heavy Fermi-liquid like down to zero temperature.

The model we consider here is rather artificial -- in particular, there are $N^4$ independent order parameters, treated as static but random variables with a vanishing average and nonvanishing variance. However, we note that even if the order parameter is not random, its effect on quantities such as Green's functions and superfluid density can still only enter via the variance $|\Delta_{4e}|^2$ at lowest order. Therefore we expect our toy model, which can be analytically solved, to be able to capture certain qualitative properties for a generic charge-$4e$ superconductor, despite the complicated structure of the order parameter. In particular, we argue that the three main results we obtained --- the gaplessness, the smallness of superfluid density, and the low-temperature pairing tendency are generic features of a charge-$4e$ superconductor.

The rest of this paper is organized as follows. In Section \ref{sec:model} we present the model for a {mean-field} charge-$4e$ superconductor, its effective action, and the corresponding saddle-point equations. In Section \ref{sec:charge-$4e$} we analyze the properties of the charge-$4e$ superconductor, including its gaplessness,  its behaviors at different temperature regimes, and its superfluid density. {In Section \ref{sec:beyond}, we argue that our findings are valid beyond mean-field for an isolated charge-$4e$ superconductor preserving total charge.}   In Section \ref{sec:pairing} we consider the potential instability of the charge-$4e$ superconudctor toward charge-$2e$ superconductivity upon lowering temperature. In Section \ref{sec:discussion} we briefly comment on the implication of our results on the stability of charge-$4e$ superconductors. 

\section{The model}
\label{sec:model}

Our Hamiltonian for a charge-$4e$ superconductor reads 
\begin{align}
    H =& H_0 + H_{int} \label{H}  \\
    H_0 =& \sum_\mathbf{k} \sum_{i=1}^N \xi_{\mathbf{k}} \Psi_{\mathbf{k} i}^\dag \Psi_{\mathbf{k} i}, \\ 
    H_{int} =& \frac{k_F^{-d}}{N^{3/2}} \int d \mathbf{r} \sum_{i<j,k<l,i<k}^N \label{V} \\
    &\Big( \mathcal{J}_{ij;kl} \Psi^\dag_{\mathbf{r} i} i \sigma_y (\Psi_{\mathbf{r}j}^\dag)^T  \,  \Psi_{\mathbf{r} k}^T i \sigma_y^T \Psi_{\mathbf{r} l}  \nonumber\\
    &+ {\Delta_{4e}}_{,ij;kl} \, \Psi^\dag_{\mathbf{r} i} i \sigma_y (\Psi_{\mathbf{r} j}^\dag)^T  \,  \Psi^\dag_{\mathbf{r} k} i \sigma_y (\Psi_{\mathbf{r} l}^\dag)^T  + h.c.\Big), \nonumber
    \end{align}
where \be \xi_{\bf k}=\frac{{\bf k}^2}{2m}-\mu \label{xi_k}\ee
{is the fermionic} dispersion with the Fermi energy 
{$\varepsilon_F = k_F^2/(2m)$}, $v_F {= k_F/m}$ is the Fermi velocity, $k_F$ is the Fermi momentum, $m$ is the fermion mass, and $d$ is the spatial dimension. 
In Eq.~\eqref{V} 
the form of interaction is similar to that in the { translationally invariant} complex SYK model~\cite{Chowdhury2018Translationally} with a charge-$4e$ quartic term~\cite{Jiang2017Charge}. Indeed, the first term in the Hamiltonian Eq.~(\ref{H}) describes the SYK interaction of $N$ flavors of spin-$1/2$ fermions $\Psi_{\mathbf{r} i} = \begin{pmatrix} \psi_{\mathbf{r} i\uparrow} & \psi_{\mathbf{r} i\downarrow} \end{pmatrix}^T$, while the second term introduces  four-fermion interactions in  forms of pair hopping ($\mathcal{J}$) and ``pair pairing" ($\Delta_{4e}$). {It is implied that $\hbar = k_B = e =1$ throughout the paper.}

As in the SYK model, we take the large-$N$ limit, in which the coupling constants $\mathcal{J}_{ij;kl}$ and ${\Delta_{4e}}_{,ij;kl}$ are constant in space and real {independent} Gaussian random variables with respect to flavor indices with finite variances 
\be
\overline{\mathcal{J}_{ij;kl}^2} = \mathcal{J}^2,~~~ \overline{{\Delta^2_{4e}}_{,ij;kl}} = \Delta_{4e}^2.
\ee
In addition, we also assume a finite correlation between $\mathcal{J}$ and $\Delta_{4e}$, such that
\be
\overline{\mathcal{J}_{ij;kl}{\Delta_{4e}}_{,ij;kl}} = \rho\mathcal{J}\Delta_{4e},
\ee
where $\rho \in (-1,1) $ sets the correlation between the two ensembles. Formally,
the correlated sets $\mathcal{J}_{ij;kl}$ and ${\Delta_{4e}}_{ij;kl}$ of random variables are described by the bivariate Gaussian distribution~\cite{Abramowitz1964Handbook}:
\begin{align}
   & P(\mathcal{J}\!,{\Delta_{4e}}\!) \! \propto \label{P_JK} \\
    & \exp\[{-\frac{1}{2(1-\rho^2)}\!\left(\!\frac{\mathcal{J}^2_{ij;kl}}{\mathcal{J}^2} + \frac{{\Delta_{4e}\!}_{,ij;kl}}{\Delta_{4e}^2} - 2\rho \,\frac{\mathcal{J}_{ij;kl} {\Delta_{4e}\!}_{,ij;kl}}{\mathcal{J} \Delta_{4e}}\!\right)}\]. \nonumber
\end{align}
For the majority of our work, we will consider the nontrivial strong-coupling limit, in which $(\mathcal{J}^2+\Delta_{4e}^2)\gg \varepsilon_F$. In the opposite weak-coupling limit, one expects a more conventional system behavior that is smoothly connected with the free Fermi gas. We will briefly comment on this regime when we discuss the superfluid density in Sec. \ref{sec:S_density}. 

Our model preserves spin-rotation symmetry, and thus we take the fermionic Green's function to be spin-diagonal:
\begin{align}
    -N^{-1}\!\sum_i^N\langle \Psi_{\mathbf{r} i}(\tau) \Psi_{0i}^\dag(0) \rangle \!=\! \sigma_0 \, G(\tau, \mathbf{r}). \label{G_def}
\end{align}
For later purposes, we also introduce an anomalous (pairing) Green's function in the equal-flavor, spin-singlet channel
\begin{align}
    {\hat{F}(\tau,{\bf r})}&\!\!=\!-N^{-1}\!\sum_i^N\langle \Psi_{\mathbf{r} i}(\tau) \Psi_{0 i}^T(0) \rangle \!\!=\! i \sigma_y \, F(\tau, \mathbf{r}), \label{F_def}\\
     {\hat{F}^+\!(\tau,{\bf r})}&\!\!=\!-N^{-1}\!\sum_i^N\langle \Psi^*_{\mathbf{r} i}(\tau) \Psi_{0 i}^\dag(0) \rangle \!\!=\! -i \sigma_y  \, F^*(\tau, \mathbf{r}). \label{F+_def}
\end{align}
We emphasize that the anomalous Green's function is not a direct consequence of the charge-$4e$ condensate, but rather of a possible transition toward a charge-$2e$ condensate. We will address this  in Section \ref{sec:pairing}.

Applying the standard machinery widely used in the SYK-like models~\cite{Patel2018Coherent,Cheipesh2019Reentrant,Esterlis2019Cooper,Hurtubise2020Superconducting} 
generalized to include potential pairing effects discussed in details in Appendix \ref{app:saddle-point}, we perform the disorder average and decouple the interactions with the bi-local fields using Lagrangian multipliers $\Sigma$, $\Phi$, $\Phi^*$. We obtain the large-$N$ effective action in imaginary time for the Hamiltonian (\ref{H})
\begin{widetext}
\begin{align} \nonumber
    -\frac{S}{N} =& \sum_{\omega,\mathbf{k}} \ln \bigg( \left(i\omega -\xi_\mathbf{k} -\Sigma(k) \right)\left(i\omega + \xi_\mathbf{k} +\Sigma(-k) \right) -  \Phi^*(k)\Phi(k)  \bigg) \\ \nonumber 
    &+ \frac{A}{T} \int_x \bigg( 2 \Sigma(x)G(-x) +  \Phi(x) F^*(x) + \Phi^*(x) F(x) + \frac{k_F^{-2d}\mathcal{J}^2}{2}\left( G(x)^2 G(-x)^2 + F^*(x)^2 F(x)^2 \right) \\  &+ \frac{k_F^{-2d}\Delta_{4e}^2}{4} \left( G(x)^4 + G(-x)^4 + F^*(x)^4 + F(x)^4\right) - \frac{k_F^{-2d} \rho \mathcal{J} \Delta_{4e}}{2}  \left(F^*(x)^2 +F(x)^2 \right)\left(G(x)^2 +G(-x)^2 \right) \!\!\bigg), \label{S}
\end{align}
\end{widetext}
where $T$ is temperature and $A$ is the system's size. Here $\sum_\omega$ denotes summation over the Matsubara frequencies {and $\int_x =  \int d\tau \int d \mathbf{r}$}. The fields $\Sigma$, $\Phi$, $\Phi^*$ and $G$, $F$, $F^*$ are bi-local, e.g., $\Sigma=\Sigma(x, x')$ (with $x = (\tau, \mathbf{r})$), and due to translational invariance it only depends on $x-x'$. We have thus defined, e.g.,  $\Sigma(x)\equiv \Sigma(x,0)$ and its Fourier transform $\Sigma(k)$ (with $k=(\omega, \mathbf{k})$), to simplify notations.

The fields $\Sigma$ and $\Phi$ enter the effective action (\ref{S}) as the self-energy and the pairing vertex. Specifically, the variation of the effective action with respect to $G$ and $F$ produces the first pair of the Schwinger-Dyson (SD) equations 
\begin{align} \nonumber 
   k_F^{2d} \, \Sigma(x) =& - \mathcal{J}^2 G(x)^2 G(-x) - \Delta_{4e}^2 G(-x)^3 \\ &+ \rho \mathcal{J} \Delta_{4e} \left( F^*(x)^2 + F(x)^2 \right) \!G(-x), \label{Sigma}\\ \nonumber 
   k_F^{2d} \, \Phi^*(x) =& - \mathcal{J}^2 F^*(x)^2 F(x) - \Delta_{4e}^2 F(x)^3 \\ &+ \rho \mathcal{J} \Delta_{4e} \left( G(x)^2 + G(-x)^2 \right) \! F(x) \label{Phi}.
\end{align}
The variation of the effective action with respect to $\Sigma$ and $\Phi$ gives the second pair of the SD equations for the full Gor'kov Green's function
\begin{align}
    G(k) \!=& \frac{i\omega \!+\! \xi_\mathbf{k} \!+\!\Sigma(-k)}{\left(i\omega\!-\!\xi_\mathbf{k}\!-\!\Sigma(k)\right)\!\left(i\omega\!+\!\xi_\mathbf{k}\!+\!\Sigma(-k)\right)\!-\!\Phi^*(k)\Phi(k)}, \label{G} \\
    F^*\!(k) \!=& \frac{\Phi^*(k)}{\left(i\omega\!-\!\xi_\mathbf{k}\!-\!\Sigma(k)\right)\!\left(i\omega\!+\!\xi_\mathbf{k}\!+\!\Sigma(-k)\right)\!-\!\Phi^*(k)\Phi(k)}. \label{F}
\end{align}

\section{Properties of the charge-$4e$ superconducting state}
\label{sec:charge-$4e$}

We begin our analysis by considering the charge-$4e$ state with  $\Phi^* = \Phi =0$. As we mentioned, the vertices $\Phi$, $\Phi^*$ in the effective action (\ref{S}) correspond to the conventional charge-$2e$ pairing, which may develop via a low-temperature instability that we consider in Section \ref{sec:pairing}.
In this regime, the effective action is structurally similar to the translationally invariant lattice model studied in Ref.~\cite{Chowdhury2018Translationally}:
\begin{widetext}
\begin{align} 
    -\frac{S}{N} =& \,2 \sum_{\omega,\mathbf{k}} \ln \! \bigg(\!\!- i\omega \!+\!\xi_\mathbf{k} \!+\! \Sigma(k)  \!\bigg) \! + \! \frac{A}{T}  \!\int_x\!\! \bigg( \!2 \Sigma(x)G(-x) \!+\! \frac{k_F^{-2d} \mathcal{J}^2}{2} G(x)^2 G(-x)^2 \!+\! \frac{k_F^{-2d} \Delta_{4e}^2}{2}  G(x)^4 \bigg), \label{S_4e}
\end{align}
\end{widetext}
with additional $U(1)$-breaking $\Delta_{4e}$-terms in the interaction Hamiltonian (\ref{V}) contribute to the effective action. 

\subsection{Heavy Fermi liquid and non-Fermi liquid behaviors}
\label{sec:FL}
We derive the Green's function of the Hamiltonian (\ref{H}) in absence of charge-$2e$ pairing.
In this case, the Schwinger-Dyson equations \eqref{Sigma} simplify to 
\begin{align}
    \Sigma(k) \!=\!& - \! T k_F^{-d} \sum_{q}\!\left( \mathcal{J}^2 \Pi(q) \!+\! \Delta_{4e}^2 \Pi(-q) \right)\!G(q-k), \label{Sigma_4e} \\ 
    \Pi(q) \!=& T k_F^{-d} \sum_k G\left(\frac{q}{2}+k\right) G\left(\frac{q}{2}-k\right), \label{PP_4e}  \\
    G(k) \!=& \frac{1}{i\omega-\xi_\mathbf{k}-\Sigma(k)}. \label{G_4e}
\end{align}
Here $\Pi(q)$ is the particle-particle bubble, and the self-energy (\ref{Sigma_4e}) correspond to the melonic diagrams with $\mathcal{J}$- and $\Delta_{4e}$-vertices.

The solution of the Schwinger-Dyson equations (\ref{Sigma_4e}-\ref{G_4e}) is quite similar to that for the 2d model considered in Ref.~\cite{Chowdhury2018Translationally}. In the strong-coupling limit, the bandwidth of the free fermions $\sim \varepsilon_F$ and the characteristic strength of the SYK-like interaction 
\be
U =\sqrt{\mathcal{J}^2 + \Delta_{4e}^2}
\ee
give rise to an intermediate energy scale $ \varepsilon_F^2/U$, which corresponds to the renormalized bandwidth.

Within the renormalized bandwidth, that is, $\omega, T \ll \varepsilon_F^2/U$, the system behaves as a heavy Fermi liquid. The self-energy is given by 
\be
\Sigma(\omega,{\bf k}) = -i Z^{-1} \omega + \frac{\beta}{4} \, {\bf v}_F\cdot\mathbf{k},
\label{eq:beta1}
\ee
where $Z\sim \varepsilon_F/U\ll 1$ and $0<\beta  = \mathcal{O}(1)$. Note that different from the result in Ref.~\cite{Chowdhury2018Translationally}, we find that generically $\beta>0$; this will be important to the result of superfluid density. { We introduce the factor of four in the equation \eqref{eq:beta1} to simplify the expression for the superfluid density in Section \ref{sec:S_density}.} The Green's function has a quasiparticle form
\begin{align}
    G(\omega, \mathbf{k}) = \frac{Z}{i\omega -\mathbf{v}_F^*\cdot\mathbf{k}}, \quad \omega, T \ll \varepsilon_F^2/U, \label{G_QP}
\end{align}
where 
$v_F^*  \sim  Z v_F$ is the renormalized Fermi velocity. 
We leave the detailed derivation of the self-energy and Green's function for Appendix \ref{app:B1}.

At higher energies above the renormalized bandwidth, $\varepsilon_F^2/U\ll \omega, T \ll U$, the system behavior is essentially local~\cite{Chowdhury2018Translationally}. We thus expect the Green's function at leading order to be the same as that for the zero-dimensional SYK model:
\begin{align}
   \Sigma(\omega, \mathbf{k})\simeq -i \pi^{-1/4} {\sqrt{U|\omega|}} {\mathrm{sgn}(\omega)}.
\end{align} 
In this regime, since $\Sigma\gg i\omega-\xi_{\mathbf{k}}$, the Green's function is given by 
\begin{align}
   G(\omega, \mathbf{k})\simeq   -i \pi^{1/4}\frac{\mathrm{sgn}(\omega)} {\sqrt{U|\omega|}}, \quad \varepsilon_F^2/U\ll \omega, T \ll U. \label{G_SYK}
\end{align}

Indeed, one can verify that the momentum dependence of $\Sigma(\omega,
\mathbf{k})$ is parametrically small, suppressed by $\varepsilon_F/U$, i.e.
\be
  \Sigma(\omega, \mathbf{k})\simeq -i \pi^{-1/4} {\sqrt{U|\omega|}} {\mathrm{sgn}(\omega)} + \frac{\beta}{4}  \,{\bf v}_F \cdot\mathbf{k},
  \label{eq:beta2}
\ee
where $\beta = \mathcal{O}(\varepsilon_F/U)$.
This suppression can be understood as follows. As pointed out in Ref.~\cite{Chowdhury2018Translationally}, for $T\gg \varepsilon_F^2/U$ the $\bf k$-dependent terms in the Green's fucntion is small compared with $\omega$-dependent terms and can be treated perturbatively in the small parameter $\varepsilon_F/U$. To obtain the $\bf k$-dependence in $\Sigma(\omega, \mathbf{k})$, at leading order, one keeps $\bf k$-dependence in only one of the Green's functions. At this order, after integrating over momentum, the resulting contribution is completely $\bf k$-independent. Therefore, the leading $\bf k$-dependent contribution to $\Sigma(\omega, \mathbf{k})$ is suppressed by $\varepsilon_F/U$. We show more details in Appendix \ref{app:high_T}.

Before we end this subsection, we briefly mention that for $T\gg U$, the system behaves as a weakly-interacting Fermi gas, just like the regular SYK model~\cite{Sachdev1993Gapless}. We also note that so far we have focused on the strong-coupling regime with a nontrivial intermediate energy scale. In the oppsite weak-coupling limit $\varepsilon_F\ll U$, we expect the ground state is a Fermi liquid perturbatively connected to free fermions. 

\subsection{Luttinger relation}

In a charge-$2e$ superconductor, the order parameter couples bilinearly to the fermion operators, which scatter an electron into a hole. Through a Bogoliubov transformation, one explicitly obtains the spectrum of gapped excitations, in quanta of the Bogoliubons that are linear superposition of an electrons and a hole~\cite{Bogoljubov1958New,Altland2010Condensed}. However, for a charge-$4e$ superconudctor, the order parameter behaves as an anomalous four-fermion interaction. As we have seen, the ground state is gapless with a Fermi surface and in this sense similar to a Fermi liquid.  A natural question is whether the volume enclosed by the Fermi surface is related to the (average) number density of the system, which is true for a real Fermi liquid by the celebrated Luttinger's theorem.

Historically, for a Fermi liquid, Luttinger's theorem was proven perturbatively via the Luttinger-Ward functional~\cite{Luttinger1960Fermi,Luttinger1960Ground}. Later, it was proven topologically~\cite{Oshikawa2000Topological, Else2021Non-Fermi} by directly connecting the ultraviolet (UV) theory with its infrared (IR) properties with a 't Hooft anomaly. The proofs in the second category explicitly requires a $U(1)$ symmetry, which our model lacks. Nevertheless, it is still interesting to investigate the fate of Luttinger's theorem for a charge-$4e$ superconductor via the Luttinger-Ward functional.

The Luttinger-Ward (LW) functional $\Phi[G]$ is defined via the relation~\cite{Luttinger1960Ground}:
\begin{align}
    \Sigma(\omega,\mathbf{k})= \frac{\delta \Phi[G(\omega,\mathbf{k})]}{\delta G(\omega,\mathbf{k})}.
\end{align}
For a regular Fermi liquid, the Luttinger-Ward functional was contructed via summing over two-particle irreducible vacuum diagrams order by order~\cite{Luttinger1960Ground}. In our SYK-like model, the random all-to-all nature of the interactions in SYK-like models~\cite{Georges2001Quantum,Chowdhury2018Translationally} allows us to explicitly obtain the Luttinger-Ward functional.

For the SYK-like models, it is straightforward to fulfill this requirement because of the structure of the self-energy equation~\cite{Georges2001Quantum,Chowdhury2018Translationally}. Indeed, from (\ref{Sigma_4e}) with $F=0$, one has
\begin{align}
\Phi[G]\!=\! -{k_F^{-2d}}\!\!\int_{x} \! \(\mathcal{J}^2G(x)^2G(-x)^2\!+\! \Delta_{4e}^2  G(-x)^4\)\!. \label{LW}
\end{align}
We note that unlike a charge-$2e$ superconductor whose Luttinger-Ward functional is not single-valued~\cite{Altshuler1998Luttinger}, here $\Phi[G]$ is a well-behaved functional of $G$. 

To check the Luttinger relation, the particle density (per flavor) reads 
\begin{align}
    n \!=\! 2 \!\sum_{\mathbf{k}}G(\tau\!=\!0^+\!\!, \mathbf{k}) \!=\! 2\!\!\int \!\! \frac{d\omega}{2\pi} \! \int \!\!\frac{d\mathbf{k}}{(2\pi)^2} G(\omega,\mathbf{k})  e^{\!-i \omega 0^+} \label{n_4e}
\end{align}
with a factor of two originating from spin. Using the equation for the Green's function (\ref{G}), we substitute 
\be
{i}G(\omega, \mathbf{k})\!=\!  G(\omega, \mathbf{k})\partial_\omega G(\omega, \mathbf{k})^{\!-1} \!+\! G(\omega, \mathbf{k}) \partial_\omega \Sigma (\omega, \mathbf{k}) \label{iG}
\ee
to Eq.~(\ref{n_4e}). 
When integrated over $\omega$, the first term is given by
\begin{align} \nonumber
I_1=& 2{i} \int \!\!\frac{d\mathbf{k}}{(2\pi)^2} \int \!\! \frac{d\omega}{2\pi} \partial_\omega \ln G(\omega,\mathbf{k}) e^{-i \omega 0^+} \\ \nonumber  =& - \frac{1}{\pi} \!\int \!\!\frac{d\mathbf{k}}{(2\pi)^2} \left( \arg(G^R(0, {\bf k})) \!-\! \arg(G^A(0, {\bf k})) \right) \\=& 2 \int_{|\mathbf{k}|\leq k_F}\frac{d\mathbf{k}}{(2\pi)^2},
\end{align}
which is the volume of the Fermi liquid.
Here we have deformed the contour to the imaginary axis, and used the analytic properties of the Green's functions $G(\omega \to -i\omega \pm 0^+) =  G^{R,A}(\omega)$ as well as the stability of the ground state enforcing $\Imm \Sigma(0,\mathbf{k})=0$.

When integrated over $\omega$, the second term  in Eq. (\ref{iG}) is given by
\begin{align}
I_2=    
-\int_\omega \Sigma(\omega, \mathbf{k}) \partial_\omega G (\omega, \mathbf{k}) = -\int_\omega \frac{\delta \Phi}{\delta G} \partial_\omega G,
\end{align}
Here it's tempting to claim that, using a chain rule the integrand is a total derivative ``$\partial \Phi(\omega)/\partial\omega$", and thus $I_2=\int_\omega \partial_\omega \Phi(\omega)=0$. {However, this is not true ---} from the definition of the functional derivative, we have
 \begin{align} \nonumber 
 I_2
 =&\!-\!\left.{\frac {d}{d\varepsilon }}\Phi[G(\omega,k) \!+\!\varepsilon \partial_\omega G(\omega,k) ]\right|_{\varepsilon =0} \\ =&\!-\!\left.{\frac {d}{d\varepsilon }}\Phi[G (\omega+\epsilon,k) ]\right|_{\varepsilon =0}\! \!\!\!\!\!\!=\!-\!\left.{\frac {d}{d\varepsilon }}\Phi[G(x)e^{\!-i\varepsilon\tau} ]\right|_{\varepsilon =0}\!.
  \label{eq:12}
  \end{align}
For a regular Fermi liquid, this vanishes due to the $U(1)$ symmetry. However, it is clear that the $\Delta_{4e}^2$ term in Eq.~\eqref{LW} is not invariant under $G(x) \to G(x) e^{i\varepsilon \tau}$, due to the broken $U(1)$ symmetry of a charge-$4e$ superconductor. Therefore, $I_2\neq 0$, and the usual Luttinger relation between the average number density and the Fermi surface volume is violated~\footnote{Note that this conclusion is contrary to the result in a previous version of this paper.}.
  
The evaluation of $I_2$ involves UV details of the theory, and we leave the derivation of a modified Luttinger relation to a future work.



\subsection{Superfluid density} \label{sec:S_density}

For a  system of $N$ fermionic flavors, the superfluid density $n_s$ (per flavor) is given by the photon mass term generated by integrating out the fermions:
\be
\mathcal{L}_{\bf A} =  {N}\frac{n_s}{2m}{\bf A}^2.
\ee
Using the parabolic dispersion $\xi_{\bf k}=({\bf k}-e{\bf A})^2/2m-\mu$, there are two contributions to $n_s$ from the fermions, known as diamagnetic and paramagnetic terms \cite{Altland2010Condensed}:
\be
n_s = n + m\Pi(0),
\label{eq:ns}
\ee
where $\Pi(0)$ is the current-current correlator. For an isotropic system, without loss of generality,
\be
\Pi(0) = \int_k \frac{{{\bf k}}}{m} { \cdot} { {\bf\Gamma}}(k,k)G(k)^2
\label{eq:jj}
\ee
where $k=(\omega, {\bf k})$ and $ { {\bf\Gamma}}(k, k)$ is the renormalized current vertex whose bare value is $ {\bf k}/m$.

In a normal metal, these two terms exactly cancel, which is a direct consequence of the  Ward identity for $p=(0,\mathbf{p})$
\begin{align}
    {{\bf p} \cdot {\bf \Gamma}}(k,k+p)  G(k) G(k+p) =&  - G(k) + G(k+p),
    \label{eq:ward}
\end{align}
or in the  $\mathbf{p}\to 0$ limit,
\be
    \mathbf{\Gamma}(k,k)  =  - \partial_{\mathbf k} G^{-1}(k).
    \label{eq:34}
\ee
Indeed, plugging \eqref{eq:34} to \eqref{eq:jj} and integrating by parts, we find $n_s=0$ in \eqref{eq:ns}. 
For completeness, we prove the Ward identity in Appendix \ref{app:ward_metal}.

The Ward identity is, in turn, a direct consequence of the particle number conservation. In a superconductor with no particle number conservation, the original Ward identity Eq.~\eqref{eq:ward} is violated. Instead, we show in Appendix \ref{app:ward_4e} for a charge-$4e$ superconductor, there are two additional terms in the modified Ward identity, which are Fourier transforms of the six-point functions
\begin{align}
4 \Delta_{4e} \!\langle \psi(x_1) \bar{\psi}(x_2) \bar\psi^4(x)\rangle \!-\!4  \Delta^*_{4e} \!\langle \psi(x_1) \bar{\psi}(x_2) \psi^4(x)\rangle ,
 \label{eq:20}
\end{align}
where symbolically $\psi^4(x)$ represents the quartetting term for the charge-$4e$ superconductor that is compatible with fermionic statistics. 
 
\begin{figure*}[t!!]
\center
\includegraphics[width=0.844\linewidth]{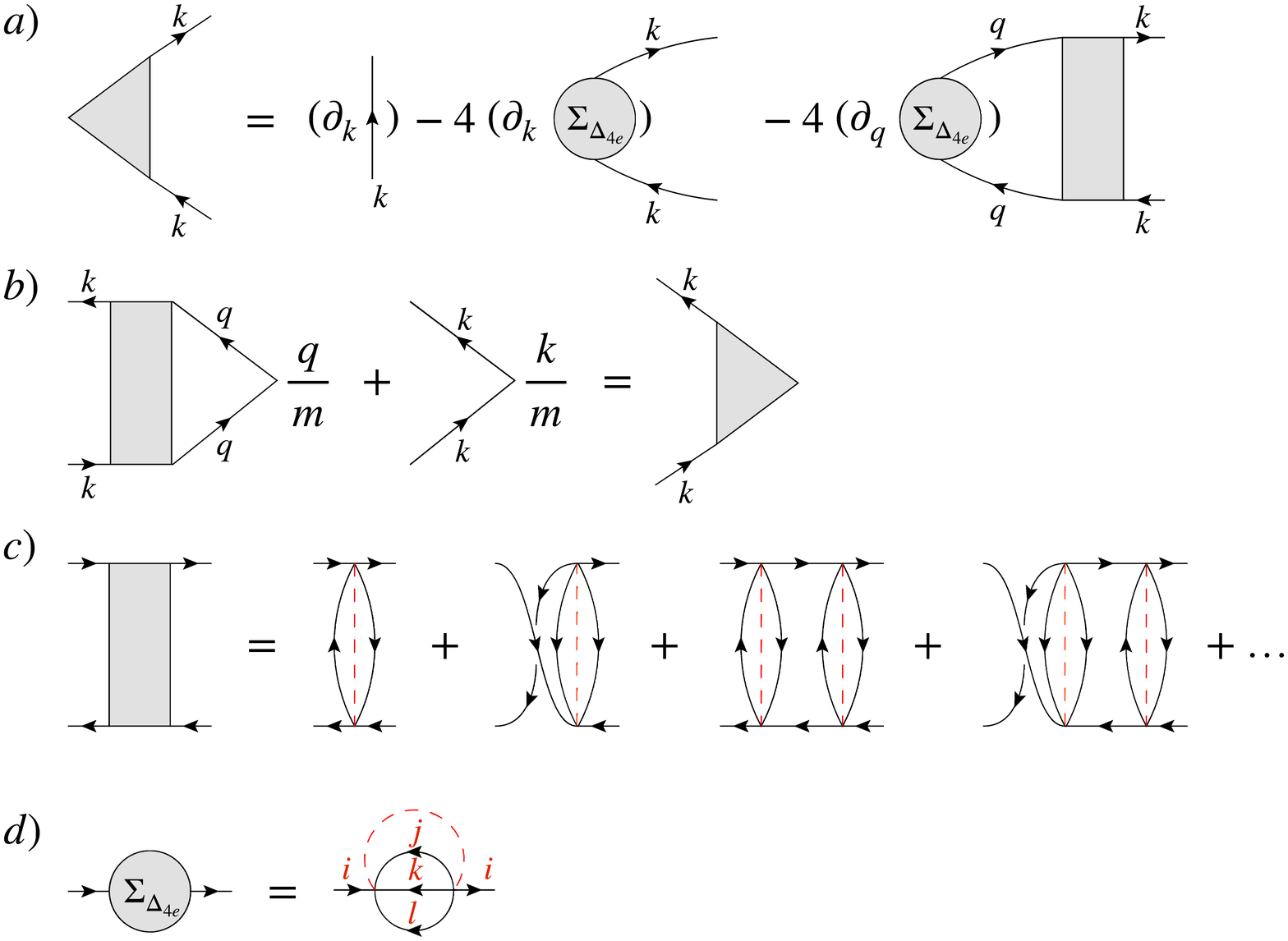} 
\caption{\small \label{fig:diagrams} {Panel (a): The diagrammatic representation of the modified Ward identity for the charge-$4e$ superconductor \eqref{eq:38}.  The gray triangle denotes the current vertex $(\mathbf{k}/m) \mathbf{\Gamma}(k, k)$, while the gray rectangle designates the four-point vertex $\chi(q,q,k,k)$. The solid black lines are the renormalized Green's functions and the dashed red lines are the disorder contractions. Panel (b): Fully renormalized current vertex \eqref{eq_vertex}. Panel (c): The four-point function $\chi(q,q,k,k)$ of the charge-$4e$ superconductor. Panel (d): The self-energy due to the ``pair-pairing'' interaction.}}
\end{figure*} 
 
For a general interacting model, the six-point functions in Eq.~\eqref{eq:20} involves infinite numbers of diagrams and cannot be expressed in a closed form. However, in our SYK-like model, we only need to include melonic diagrams to leading order in $1/N$, which enables us to obtain analytically
\begin{widetext}
\begin{align}
    &{{\bf p} \cdot {\bf \Gamma}}(k,k+p)  G(k) G(k+p)  
    =    { -} G(k) { +} G(k+p) \nonumber + 4\Sigma_{\Delta_{4e}}(k) G(k)G(k+p) -  4\Sigma_{\Delta_{4e}}(k+p) G(k)G(k+p) \nonumber \\
    & +4\int_q \Sigma_{\Delta_{4e}}(q) G(q)G(q+p) \chi(q,q+p,k,k+p)G(k)G(k+p)\nonumber \\
    &- 4\int_q \Sigma_{\Delta_{4e}}(q+p) G(q)G(q+p) \chi(q,q+p,k,k+p)G(k)G(k+p),
    \label{eq:21}
\end{align}
\end{widetext}
{as shown in Fig. \ref{fig:diagrams} for the case of ${\bf p} \to 0$.} 
Here $\Sigma_{\Delta_{4e}}$ is a self-energy-like diagram but with only $\Delta_{4e}$ vertices (see Fig.~\ref{fig:diagrams}(d)), which satisfies in our model 
\be
\Sigma_{\Delta_{4e}}(\omega, \mathbf{k}) = \frac{\Delta_{4e}^2}{\Delta_{4e}^2+\mathcal{J}^2}\Sigma(\omega, \mathbf{k}).
\label{eq:sigma4e}
\ee
The four-point vertex $\chi(q,q+p,k,k+p)$ is a given by the sum of a series of ladder diagrams. 
Within our SYK-like model, at each rung there are four types of diagrams with $\mathcal{J}$ and $\Delta_{4e}$ vertices. The four-point vertex $\chi$ and the fully renormalized current vertex $\Gamma$ are related by
\be
\mathbf{\Gamma}(k,k) = \frac{\bf k}{m}+ \int_q \chi(k,k,q,q)G^2(q)\frac{{\bf q}}{m}.
\label{eq_vertex}
\ee

{Taking the ${\bf p}\to 0$ limit in Eq. (\ref{eq:21})}, we have 
\begin{align} \nonumber 
&\mathbf\Gamma(k,k) {G(k)^2} = \partial_{\bf k} G(k) -4\partial_{\bf k}\Sigma_{\Delta_{4e}}(k)G^2(k)\\&- 4\hat {\bf k} \int_q 
\hat{\bf k}\cdot\partial_{\bf q}\Sigma_{\Delta_{4e}}(q)G^2(q)\chi(q,q,k,k)G^2(k),
\label{eq:38}
\end{align}
{where $\hat {\bf k}$ is a unit vector in the direction of the internal momentum ${\bf k}$.}
Compared with \eqref{eq:34}, the additional two terms here represent the effect of non-conservation of particle number and are responsible for the superfluid density $n_s$. We illustrate the modified Ward identity diagrammatically in Fig.~\eqref{fig:diagrams}. 

Multiplying both sides by $k$ and integrating over $k$ we obtain the paramagnetic contribution to  $n_s$ in Eq.~\eqref{eq:ns},
\begin{align}
& m\Pi(0)=\int_k {\bf k} \cdot {\bf \Gamma}(k,k)G^2(k) \nonumber \\ &= \int_k {\bf k} \cdot \partial_{\bf k} G(k)  -4\int_k  {\bf k} \cdot \partial_{\bf k} \Sigma_{\Delta_{4e}}(k)G^2(k)  \nonumber \\ &- 4\int_k\int_q {\bf k} \cdot \partial_{\bf q}\Sigma_{\Delta_{4e}}(q)G^2(q)\chi(q,q,k,k)G^2(k).
\end{align}

Next, swapping $k$ and $q$ in the last term, and using the relation \eqref{eq_vertex} between the four-point vertex $\chi$ and the current vertex $\Gamma$, we obtain
\begin{align}
&\int_k {\bf k} \cdot {\bf \Gamma}(k,k) G^2(k) \nonumber \\ &= \int_k {\bf k} \cdot  \partial_{\bf k} G(k) -4m\int_k \partial_{\bf k} \Sigma_{\Delta_{4e}}(k) {\bf \Gamma}(k,k) G^2(k), \label{eq:ward4e}
\end{align}
and thus
\begin{align}
& \int_k \({\bf k}+4m \partial_{\bf k}\Sigma_{\Delta_{4e}}(k)\) {\bf \Gamma}(k,k) G^2(k) \nonumber \\&=  \int_k {\bf k} \cdot \partial_{\bf k} G(k) = -n.
\label{eq:44b}
\end{align}

Assuming the main contribution to the left hand side  of Eq.~\eqref{eq:44b} comes from the Fermi surface, we have
\begin{align}
m\Pi(0)&=\int_k {\bf k} \cdot {\bf \Gamma}(k,k) G^2(k) \nonumber \\ &\simeq -\frac{n}{1+\dfrac{4}{v_F}\, \left.\hat{\bf k} \cdot\partial_{\bf k} \Sigma_{\Delta_{4e}}(k)\right|_{{\bf k}={\bf k}_F}},
\end{align}
and the superfluid density is
\be
n_s  = n + m\Pi(0) = \frac{n\alpha }{1+\alpha},
\ee
where 
\be
\alpha = \frac{4}{v_F}\, \left.{\hat{\bf k} \cdot \partial_{\bf k}}\Sigma_{\Delta_{4e}}(k)\right|_{{\bf k}={\bf k}_F}.
\ee

In the present model, using Eq.~\eqref{eq:sigma4e} and Eqs.~(\ref{eq:beta1}) we have
\be
\alpha=
{\beta}\frac{\Delta_{4e}^2}{\mathcal{J}^2+\Delta_{4e}^2},
\label{eq:29}
\ee
where we remind that $\beta\equiv4\,\hat{\bf k} \cdot\partial_{\bf k} \Sigma(k)|_{{\bf k}={\bf k}_F}/v_F$ is a non-universal $\mathcal{O}(1)$ constant in the heavy FL regime,
and thus
\be
n_s = \frac{\beta \Delta_{4e}^2}{\mathcal{J}^2+ (\beta+1)\Delta_{4e}^2} \, n.
\label{eq:48}
\ee

Eq.~\eqref{eq:48} is one of the key results of this work. From the $f$-sum rule \cite{Mahan2000Many} and the stability conditon of the ``Higgsed" gauge field, we expect that $0\leq n_s\leq n$. As Eq.~\eqref{eq:48} remains valid at $\mathcal{J}=0$, this in turn requires $\beta>0$. 
Indeed, this is what we found for the heavy FL regime in Sec.~\ref{sec:FL}, different from the result in Ref.~\cite{Chowdhury2018Translationally}.

It is now straightforward to obtain the superfluid density of the charge-$4e$ superconductor in various regimes:
\begin{itemize}
    \item In the strong-coupling limit, for $T\ll \varepsilon_F^2/U$, the system behaves qualitatively  as its heavy Fermi-liquid ground state. As we have found in Sec.~\ref{sec:FL}, $0<\beta\sim \mathcal{O}(1)$. In this regime the superfluid ratio $n_s/n$ depends on the ratio between $\mathcal{J}$ and $\Delta_{4e}$. For $\Delta_{4e}\ll \mathcal{J}$, the superfluid density is perturbatively small in its order parameter, $n_s/n\sim \Delta_{4e}^2/\mathcal{J}^2$.  In the opposite limit including the case with $\mathcal{J}=0$, $n_s\sim n$ but the superfluid ratio is still in general smaller than one. These features are in sharp contrast with a charge-$2e$ superconductor.
    
    \item In the strong-coupling limit, for $ \varepsilon_F^2/U \ll T \ll U$, the system behaves as a non-Fermi liquid.  Strictly speaking, in this regime $T$ is much greater than the renormalized bandwidth $\varepsilon_F^2/U$, so the contribution to the momentum integral is not concentrated near the Fermi surface but is rather smeared across the entire Fermi sea. However, we expect $\partial_{\bf k}\Sigma_{\Delta_{4e}}$ in Eq.~\eqref{eq:44b} in the entire Fermi sea to be suppressed by $\varepsilon_F/U$, for the same reason why $\beta$ is so. We expect \eqref{eq:48} to remain qualitatively correct. From Sec.~\ref{sec:FL}, $\beta=\mathcal{O}(\varepsilon_F/U)$, and the superfluid density ratio is also suppressed, i.e., $n_s/n\sim \varepsilon_F/U$. 

We note in passing that in a 2d model, going beyond the mean-field theory, the charge-$4e$ order can be destroyed via a Berezinskii–Kosterlitz-Thousless transition~\cite{Berezinskii1971Destruction1,*Berezinskii1971Destruction2,Kosterlitz1973Ordering,Altland2010Condensed} due to phase fluctuations, {as it was shown in the earlier studies of charge-$4e$ superconductivity \cite{Berg2009Charge}. In our model the transition} occurs for $N n_s/n\lesssim T/\varepsilon_F$. The factor of $N$ comes from $N$ flavors of fermions, since we have assumed phase coherence of all the charge-$4e$ order parameters. Thus even though $n_s$ is small in $\varepsilon_F/U$, the quasi-long-range order remains robust in the large-$N$ limit.

    
    \item Eq.~\eqref{eq:48} remains valid for the weak coupling limit $U\ll \varepsilon_F$, in which the system is a perturbative Fermi liquid at low temperatures. In this regime, both the self-energy and $\beta$ are of the order of $(U/\varepsilon_F)^2$. Therefore $n_s/n\sim(U/\varepsilon_F)^2$.
    
\end{itemize}

We end this subsection by contrasting the superfluid density in a charge-$4e$ superconductor with that in a charge-$2e$ (BCS) superconductor. For completeness, we evaluated the superfluid density for a charge-$2e$ superconductor in Appendix \ref{app:ward_BCS} by the modified Ward identity method. Shown in  Eq.~\eqref{app:ward1}, the modified Ward identity for  the charge-$2e$ superconductor is quite similar to that for the charge-$4e$ case. In both cases the modification to the Ward identity comes from the self-energy $\Sigma_{\Delta}$ due to the superconducting order parameter. The key difference, however, is that for the charge-$2e$ order, $\Sigma_{\Delta}$ is singular even for a small $\Delta$ (see Eq.~\eqref{app:n_BCS}), leading to a much larger superfluid density $n_s=n$. We note that the same singularity in self-energy is responsible for the gapping of the Fermi surface in a charge-$2e$ superconductor. 

In this sense,  the smallness of superfluid density for a charge-$4e$  superconductor can be viewed as a  consequence of the gaplessness of the Fermi surface. As we expect the gaplessness of the Fermi surface to be a robust feature for a generic charge-$4e$ superconudctor, we expect its superfluid density ratio $n_s/n$ is generally small just like in our model (unless $\Delta_{4e}$ is much larger than any other energy scale in the problem, in which case $n_s/n$ can be a significant fraction like our case). 

\section{Fixed particle number: beyond mean-field theory}
\label{sec:beyond}

In the previous Section, many of our results are associated with the spontaneous breaking of the $U(1)$ symmetry in the mean-field theory. For an isolated system that becomes superconducting, the total particle number $N$ remains a good quantum number and the $U(1)$ symmetry is intact. An interesting question is whether our previous conclusions remain valid. In this Section we answer this question in the positive by going beyond mean-field theory and explicitly constructing a $U(1)$-symmetric wave function for the charge-$4e$ superconductor.

In a conventional charge-$2e$ superconductor there is a well-known procedure~\cite{Leggett1991Concept,Lapa2020Rigorous} to restore particle number conservation, which we briefly describe before moving to charge-$4e$ superconductors. This can be done by treating the Cooper pair wave function $\Delta$ as a dynamical Hubbard-Stratonovich auxillary field, and single out its phase degree of freedom  $\Delta = |\Delta|e^{2i\theta}$ as a quantum variable. With this modification, the action is given by
\begin{align} \nonumber
{S}[\psi,\theta] =& {\int_x \sum_{\sigma=\uparrow\downarrow}} 
{\bar{\psi}_\sigma (x) (\partial_\tau - \tfrac{\partial_{\rm r}^2}{2m} - \mu) \psi_\sigma(x)} \\ &{-\int_x} \Big( {|}\Delta{|} e^{2i\theta} {\bar{\psi}_\uparrow(x) \bar{\psi}_\downarrow(x)} 
+h.c.\Big),
\label{eq:1}
\end{align}
and the conserved particle number should be regarded as 
\begin{align}
\hat N = \hat N_{\rm fermion} + 2 \hat N_{\rm Cooper},
\end{align}
where $\hat N_{\rm fermion}={\int d{\bf r} \sum_\sigma \psi_\sigma^\dag(\bf r) \psi_\sigma(\bf r)}$ 
is the number of  electrons, and $\hat N_{\rm Cooper}$ is the number of Cooper pairs. {The phase factor} $e^{2i\hat\theta}$ can be viewed as a Cooper pair annihilation operator and satisfies the commutation relation 
\begin{align}
[\hat N_{\rm Cooper}, e^{2i\hat \theta}] = - e^{2i\hat \theta}.
\end{align}

Neither $\hat N_{\rm fermion}$ nor $2\hat N_{\rm Cooper}$ is a  good quantum number, only their sum is. Moreover, since ${S}[\psi,\theta]$ is introduced via a Hubbard-Stratonovich auxillary field that derives from a fermion-only model ${S}_f[\Psi]$, we have 
$\langle \hat N_{\rm fermion}\rangle_{S} =\langle \hat N_{\rm fermion}\rangle_{S_f}= N$ and $\langle \hat N_{\rm Cooper} \rangle =0$.  The fact that $\langle \hat N_{\rm Cooper} \rangle =0 $ can also be seen from adding a small kinetic term ${\propto (\partial_\tau} \theta)^2$ to the action and canonically quantizing the action.

After integrating out the fermions {in Eq.~(\ref{eq:1})}, one ends up with a $XY$-model action for $\theta$ 
\begin{align}
 S[\theta] = \frac{\chi}{2} (\partial_\tau \theta)^2 + \frac{n_s}{2m} ({\partial_{\bf r}} \theta)^2,
\label{eq:54}
\end{align}
where due to U(1) gauge invariance, $\chi$ is the compressibility of the system, and $n_s$ is the superfluid density.

With a fixed $N$, $\theta$ is not a good quantum number. Rather, the ordered state of $\theta$ should be thought of as exhibiting \emph{off-diagonal long-range order} (ODLRO)~\cite{Yang1962Concept}, in which 
\begin{align}
\lim_{ {x} \to \infty}\langle e^{2i \theta(x)}\, e^{-2i\theta(0)}\rangle =1.
\end{align}
With ODLRO one can neglect the fluctuation effects of $\theta$, at least in the long-distance limit. 

Accordingly, the ground state wave function of a conventional superconductor described by Eq.~\eqref{eq:1} can be modified to a number-conserving one with $N$ particles, in which $\theta$ has long range correlation but is not a good quantum number. Since $N$ and $\theta$ are conjugate variables, this can be done via a Fourier transform
\begin{align}
|N\rangle = \int d\theta  e^{-iN\theta} |{\rm BCS_\theta}\rangle\otimes  |\theta\rangle, \label{wf_N}
\end{align}
where
\begin{align}
 |{\rm BCS_\theta}\rangle=\prod_{\mathbf k} \left[u_{\mathbf k} + v_{\mathbf k} e^{2i\theta} {\psi}^\dag_{\uparrow}(\mathbf k) {\psi}^\dag_{\downarrow} (-\mathbf k)\right ] |0\rangle.
\end{align}
In the above $|0\rangle$ is the empty state for the electrons, $|\theta\rangle$ is an eigenstate for $\theta$ satisfying $\langle \theta|\theta'\rangle = \delta (\theta-\theta')$ 
~\footnote{Alternatively the wave function can be written as 
$|N\rangle = \hat P_N |0\rangle \otimes \int d\theta    |\theta\rangle$,
where
$\hat P_N= \prod_{\mathbf k} \left[u_{\mathbf k} + v_{\mathbf k} e^{2i\hat \theta} {\psi}^\dag_{\uparrow}(\mathbf k) {\psi}^\dag_{\downarrow} (-\mathbf k)\right ] e^{-iN\hat\theta}$,
with $\hat\theta |\theta\rangle = \theta |\theta\rangle$. Here $\int d\theta    |\theta\rangle$ is the vacuum for Cooper pairs, and $\hat P_N$ creates $N$ particles distributed among fermions and Cooper pairs.}. 
Similar to the coupled superfluid setup considered in Ref.~\cite{Leggett1991Concept}, here $|N\rangle$ can be thought of as a mean-field BCS state coupled with a superfluid of Cooper pairs. 
 
The important insight is that, apart from particle-number conservation, all fermionic spectral properties of the states $|N\rangle$ and $|{\rm BCS}_{{\theta}}\rangle$ are identical. Indeed, the anomalous self-energy and energy gap of a conventional self-energy only depends on $|\Delta|^2$, insensitive to averaging over $\theta$ in $|N\rangle$.

We argue that a similar prescription can be directly extended to an isolated, number-conserving charge-$4e$ superconductor. Namely, the action is given by
\begin{align}
{S}[\Psi,\theta] { = \! \int_x \! \Big( \sum_{i=1}^N \Psi_{i x}^\dag (\partial_\tau - \tfrac{\partial_{\rm r}^2}{2m} - \mu) \Psi_{i x} + H_{int}(\theta)\! \Big),}
\end{align}
{where the interaction Hamiltonian~(\ref{V}) is replaced by $H_{int}(\theta): {\Delta_{4e}}_{,ij;kl} \to {\Delta_{4e}}_{,ij;kl}\, e^{4i \theta}$. In the above, we assumed phase coherence in the flavour space.}
The conserved particle number can be written as
\begin{align}
\hat N = \hat N_{\rm fermion} + 4 \hat N_{\rm quartet},
\label{eq:8}
\end{align}
with $[\hat N_{\rm quartet}, e^{4i\hat\theta}] = - e^{4i\hat\theta}$.

We are similarly led to a quantum $XY$-model for $\theta$, whose action is of the same form as Eq.~\eqref{eq:54}. Instead of acquiring a mean-field value for $\theta$, an isolated charge-$4e$ superconductor exhibits ODLRO in $e^{4i\theta}$. With ODLRO, the ground state wave function can then be written as 
\begin{align}
|N\rangle = \int d\theta \,|{\rm MF}_\theta\rangle \otimes e^{-iN\theta} |\theta\rangle.
\label{N2}
\end{align}
Here $|{\rm MF}_\theta\rangle$ is the mean-field wave function with an order parameter at a fixed phase $\theta$. Unlike the charge-$2e$ case, the charge-$4e$ SC is an interacting system and its mean-field wave function cannot be analytically expressed.  However, due to the auxillary nature of $|\Delta|e^{4i\theta}$, we do know that
\begin{align}
N = \langle N| \hat N_{\rm fermion}|N\rangle=\langle {\rm MF}_\theta|\hat N_{\rm fermion}|{\rm MF}_\theta\rangle,
\label{N1}
\end{align}
where the second equality follows from direct evaluation on Eq.~\eqref{N2}.

In two spatial dimensions and above, at $T=0$ the ODLRO in an $XY$-model is guaranteed as long as the superfluid density $n_s>0$, which can be self-consistently verified from the fermionic sector. Remarkably, the superfluid density for an isolated charge-$4e$ superconductor is the same as that in the mean-field theory. For every $|{\rm MF}_\theta\rangle$, the superfluid density is a good quantum number and is independent of $\theta$. Indeed, as we showed in Section \ref{sec:S_density}, $n_s$ for a mean-field state is expressed via the anomalous self-energy $\Sigma_{\Delta_{4e}}$, which in turn only depends on $|\Delta_{4e}|^2$. Thus the averaging over $|\theta\rangle$ in the number-conserving state $|N \rangle$ leads to the same value of $n_s$ obtained using a mean-field theory, which is indeed positive. We see that this property obtained from mean-field theory ensures  long-range order even when fluctuation effects are included.

The superfluid density we obtained in the previous section is expressed in terms of $\langle n\rangle_{\rm MF}$ in a mean-field state, and using Eq.~\eqref{N1}, for a number-conserving system it can be equally expressed via the conserved number density $n=N/V$.

The same argument can be straightforwardly applied to show that our conclusion on the Luttinger relation extends to a number-conserving system.

Our conclusion that the charge-$4e$ SC is gapless can also be extended beyond mean-field theory. To this end, one can construct a number-conserving excited state $|N,{\rm ex}\rangle$ via the same strategy as $|N\rangle$:
\begin{align}
|N,{\rm ex}\rangle =  \int d\theta \,|{\rm MF}_\theta, {\rm ex}\rangle \otimes e^{-iN\theta} |\theta\rangle.
\end{align}
Since neither the energy of $|{\rm MF}_\theta\rangle$ nor that of $|{\rm MF}_\theta, {\rm ex}\rangle$ depends on the value of $\theta$, the energy gap of a  system  with a fixed particle number is the same as that in the mean-field theory, which vanishes in the thermodynamic limit.

\section{Pairing instability}
\label{sec:pairing}

In this section, we analyze the low-temperature instability against charge-$2e$ superconductivity. Through this low-temperature instability the system enters a low-entropy state by gapping out the Fermi surface, similar to the pairing instability in a regular Fermi liquid. The specific channel of the pairing order depends on the microscopic model; however, a similar transition may be expected for a generic charge-$4e$ superconductor~\cite{Jiang2017Charge} as long as there is an attractive interaction in certain pairing channels.

To analyze the pairing instability, we reinstate and cross-correlation $\rho$ between $\mathcal{J}$ and ${\Delta}_{4e}$, and the pairing vertex $\Phi$ in the Schwinger-Dyson equations, Eqs.~(\ref{Sigma}-\ref{F}).
The SD equations for the self-energy and for the pairing vertex are nonlinear and, thus, complicated to solve in general case. Yet, to examine the onset of pairing instability one can analyze the pairing susceptibility in the normal state, with
\begin{align}
    G_0(k) =& \frac{1}{i\omega - \xi_\mathbf{k} - \Sigma_0(k)}, \label{G0}\\
    k_F^{2d} \Sigma_0(x) =& -  \mathcal{J}^2 G_0(x)^2 G_0(-x) - \Delta_{4e}^2 G_0(-x)^3. \label{Sigma0} 
\end{align}
The self-consistent equation \eqref{Phi} for the pairing vertex $\Phi$   is given by, up to cubic order in $\Phi$
\begin{align}
   F(k) \simeq& - G_0(k) \Phi(k) G_0(-k), \label{F_pert}\\
    k_F^{2d} \Phi^*(x) \simeq& \, \rho \mathcal{J} \Delta_{4e} \left( G_0(x)^2 + G_0(-x)^2 \right) F(x) \nonumber \\&- \mathcal{J}^2 F^*(x)^2 F(x) - \Delta_{4e}^2 F(x)^3. \label{Phi_pert}
\end{align}
Note that in the presence of a nonzero $\Phi$,  the Green's function $G$ in Eq.~\eqref{G} gets renormalized by $\mathcal{O}(|\Phi|^2)$, which also leads to contribution to the pairing equation that is cubic in  $\Phi$ and proportional to $\rho \mathcal{J}{\Delta_{4e}}$. As the pairing problem is only analytically tractable at $\rho\ll 1$, we neglect this contribution.

From the linear term in Eq.~\eqref{Phi_pert} we see that effectively, the attractive pairing interaction is formed by a combination of a $\Delta_{4e}$ interaction and a $\mathcal{J}$ interaction vertices, which we show in Fig.~\eqref{fig:pairing} Since both $\Delta_{4e}$ and $\mathcal{J}$ are random variables within the flavor sector, cross correlations between the two are needed.
Indeed, we have verified that up to leading orders of $1/N$, the effective interactions with only $\mathcal{J}$ and $\Delta_{4e}$ vertices do not contribute to the pairing channel we consider.

The signs of the cubic terms in Eq.~\eqref{Phi_pert} indicate that the pairing transition is first-order~\cite{Patel2018Coherent, Hurtubise2020Superconducting}. To see this more clearly, we note that
(\ref{Phi_pert}) can be thought of as a variation of the Ginzburg-Landau (GL) functional~\cite{Landau1965Collected,Altland2010Condensed} $S_{GL} = - (NA/T) \int_x \mathcal{L}_{GL}[\Phi,\Phi^*]$ with the Lagrangian density
\begin{align} \nonumber 
   & \mathcal{L}_{GL}[\Phi,\Phi^*] \!= -\Phi(x)F^*(x) - \Phi^*(x)F(x) \\ \nonumber &+ \frac{k_F^{-2d} \rho \mathcal{J} \Delta_{4e}}{2} \!\left( G_0(x)^2 + G_0(-x)^2 \right) \left(F^*(x)^2 + F(x)^2\right) \\&- \frac{k_F^{-2d}\mathcal{J}^2}{2} |F(x)|^4 - \frac{k_F^{-2d}\Delta_{4e}^2}{4}\left( F^*(x)^4 + F(x)^4\right), \label{GL_fun}
\end{align}
where $F(k) = - G_0(k) \Phi(k) G_0(-k)$. The GL functional (\ref{GL_fun}) is a perturbative expression where the higher order terms, $F^6$, $F^8$ and so on, originate from the expansion of the 
logarithm in the effective action (\ref{S}). In absence of correlation ($\rho=0$) between the pair hopping and pair pairing terms in the Hamiltonian (\ref{H}), the GL functional (\ref{GL_fun}) has a form $\mathcal{L}_{GL} \sim a\Phi^2 - b \Phi^4 + c \Phi^6$, with $a,b,c>0$. We found that the pairing problem is quite similar to that of the original {spinless/spin-polarized} complex SYK model \cite{Sachdev2015Bekenstein,Patel2018Coherent}, which is known to not host a pairing transition \cite{Hurtubise2020Superconducting}. 
However, turning on weak positive cross correlations $|\rho| \ll 1$ changes  the GL functional in the quadratic term to $\mathcal{L}_{GL} = [a-
|\rho| \mathcal{J}\Delta_{4e}a'(T)]\Phi^2 - b \Phi^4 + c \Phi^6$, which always has two local minima for $a-|\rho| \mathcal{J}\Delta_{4e}a'(T)\to 0_+$, one at $\Phi=0$ and the other at $\Phi\neq 0$, with the latter being the global minimum. The fact that two local mininum develops before the quadratic term becomes negative indicates that the system has gone through a first-order phase transition. Therefore, at the transition temperature $T_{2e}$, we expect $\rho \mathcal{J}\Delta_{4e}a'(T_{2e})$ to be smaller than, but of the same order as $a$. In other words, at the first-order transition toward a charge-$2e$ superconductor, the eigenvalue $\lambda$ of the linearized version of Eqs.~(\ref{F_pert},\ref{Phi_pert}):
\begin{align} \nonumber 
   &\lambda \Phi^*(k) = -\rho \mathcal{J} \Delta_{4e} T k_F^{-d} \\& \times\sum_q \left(\Pi(q-k) +\Pi(k-q) \right) G_0(q) \Phi(q) G_0(-q) \label{Eliashberg}
\end{align} 
becomes of $\mathcal{O}(1)$ (but smaller than one).
Here the effective interaction is given by the particle-particle bubble
\begin{align}
    \Pi(q) = T k_F^{-d} \sum_k G_0\left(k\right)  G_0\left(q-k\right). \label{PP_pairing}
\end{align}
We note that unlike conventional linearized gap equation, the phase of $\Phi$ affects $\lambda$. This is because in a charge-$4e$ superconudctor, the $U(1)$ symmetry is broken.

\begin{figure}[t!!]
\center
\includegraphics[width=0.956\linewidth]{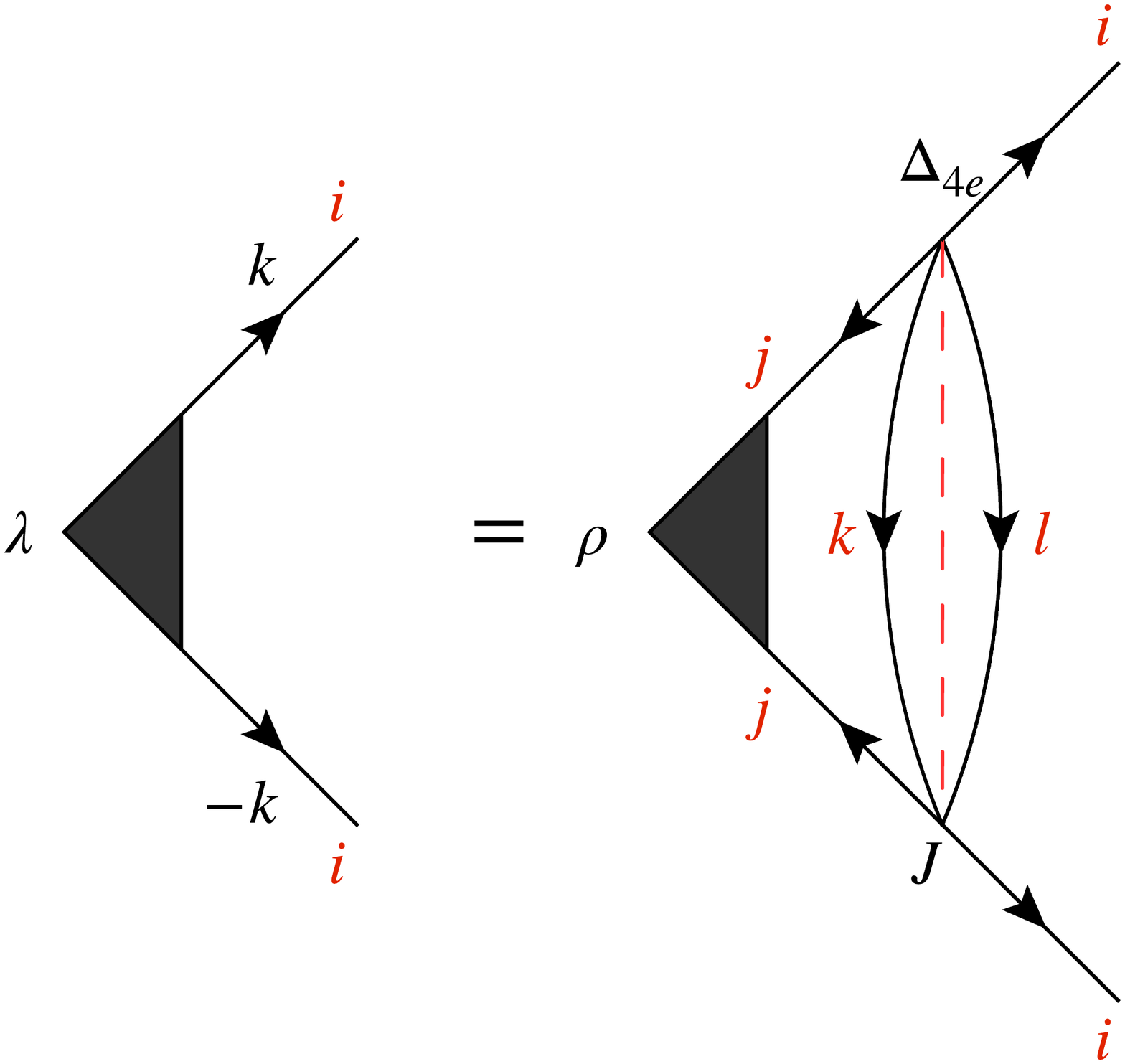} 
\caption{\small \label{fig:pairing} Diagrammatic representation of the equation for the paring vertex (\ref{Eliashberg}).
}
\end{figure}

We now analyze the eigenvalue problem \eqref{Eliashberg} for the pairing vertex in the NFL regime and in the heavy FL regime respectively. We first start from the NFL regime at high temperatures, and study the charge-$2e$ instability. If the $T_{2e}$ we obtain is higher than the characteristic NFL temperature $\varepsilon_F^2/U$, then the pairing phase develops below this temperature, preempting the heavy Fermi liquid phase. In this case, the charge-$2e$ phase can be viewed as emerging from a (charge-$4e$) non-Fermi liquid. On the other hand, if the $T_{2e}$ obtained using NFL Green's functions is lower than $\varepsilon_F^2/U$, it means that our calculation is not self-consistent -- one needs to consider contributions from both NFL and heavy FL fermions in the pairing problem.  

In the NFL regime, the particle-particle bubble $\Pi$ in Eq.~\eqref{PP_pairing} is logarithmic in frequency transfer, similar to that in color superconductivity~\cite{Son1999Superconductivity,Chubukov2005Superconductivity}. Assuming the momentum cutoff is $\mathcal{O}(k_F)$, up to a non-universal constant we have
\begin{align} \nonumber 
    \Pi(\Omega) =& - \frac{1}{2\sqrt{\pi}U} \int_{-U}^{U} d\omega \, \frac{\mathrm{sgn}\left(\omega\right)\mathrm{sgn}\left(\Omega-\omega \right)}{\sqrt{\left|\omega\right|}\sqrt{\left|\Omega-\omega \right|}} \\ \simeq &   \frac{1}{\sqrt{\pi}U} \, \ln \frac{U}{|\Omega|}, \label{PP_SYK}
\end{align}
where we have used the Green's functions in the NFL regime, as the typical frequency transfer is of the order of $T_{2e}$, which we assume for now to be much larger than $\varepsilon_F^2/U$. 

The eigenvalue problem for the  pairing vertex is then simplified to 
\begin{align}
   \lambda \Phi^*(\omega) = - \frac{\rho \mathcal{J} \Delta_{4e}}{\pi U^2} \int_{T}^U d\Omega \, \frac{\Phi(\Omega)}{\Omega} \ln \frac{U^2}{|\Omega^2-\omega^2|}, \label{gap_eq_SYK_p}
\end{align}
where the temperature $T$ appears as the IR cut-off.
The gap equation (\ref{gap_eq_SYK_p}) can be resolved with logarithmic accuracy~\cite{Son1999Superconductivity,Chubukov2005Superconductivity,Chubukov2021Pairing}. Splitting the frequency integral into two regions: $T_{2e} \leq \Omega \leq |\omega|$ and  $|\omega| \leq \Omega \leq U$, and introducing the logarithmic coordinates $x = \ln \frac{\Lambda}{|\omega|}$ and $y = \ln \frac{\Lambda}{\Omega}$, we approximate Eq. (\ref{gap_eq_SYK_p}) with 
\begin{align}
   \Phi^*(x) = -g \bigg(  \int_0^x d y \, y\, \Phi(y) + x \int_x^{\ln \frac{\Lambda}{T_{2e}}} d y \, \Phi(y)  \bigg), \label{gap_eq_SYK_log}
\end{align}
where we defined the coupling constant 
\be
g = \frac{\rho \mathcal{J} \Delta_{4e}}{\pi \lambda U^2} = \frac{\rho \mathcal{J} \Delta_{4e}}{\pi\lambda\left(\mathcal{J}^2 +\Delta_{4e}^2\right)}.
\ee
The equation (\ref{gap_eq_SYK_log}) is equivalent to the second-order differential equation \be
\frac{d^2\Phi^*(x)}{dx^2} - g \Phi(x) = 0
\ee with the boundary conditions $\Phi^*(0)=0$ and $\Phi^*_x(\ln \frac{U}{T_{2e}})=0$. A similar equation can be derived for the conjugated pairing vertex starting from Eq. (\ref{Eliashberg}). From here we deduce that the pairing vertex is imaginary (real) for $\rho>0$ ($\rho<0$), and 
\begin{align}
    \Phi(\omega) \propto \sin \left( \sqrt{g} \ln \frac{U}{|\omega|} \right), \label{pairing_SYK}
\end{align}
while the superconducting transition temperature can be found from the temperature onset satisfying $\cos (\sqrt{g} \ln \frac{U}{T_{2e}})=0$: 
\begin{align}
    T_{2e} \sim U e^{-\frac{\pi}{2\sqrt{g}}} \sim U \exp \(  {-\frac{U}{\sqrt{|\rho|\mathcal{J}\Delta_{4e} }}}\#\), \label{T_2e_SYK}
\end{align}
where $\#$ stands for a nonuniversal $\mathcal{O}(1)$  number.
The uncertainty here comes from the fact that the phase transition is first-order, which occurs when $\lambda$ is $\mathcal{O}(1)$ but smaller than one.

The solution for $T_{2e}$  above is only self-consistent if it is larger than $\varepsilon_F^2/U$. Otherwise, potential pairing can only occur in the heavy Fermi liquid regime, which requires a separate analysis. In this case, the typical frequency transfer $\Omega$ is much smaller than the $\varepsilon_F^2/U$, and there are two contributions to the particle-particle bubble, one from the NFL regime, and the other from the heavy FL regime. The contribution from the NFL regime is similar to the previous case, except that here the effective lower cutoff of the integral is instead $\varepsilon_F^2/U$: 
\begin{align}
    \Pi_{\rm NFL}(\Omega)  \simeq   \frac{1}{2\sqrt{\pi}U} \, \ln \frac{U^2}{\varepsilon_F^2}.
\end{align}
The contribution from the heavy FL regime, on the other hand, is evaluated in Appendix \ref{app:B1} (denoted as $\widetilde{\Pi}(\omega,q)$; see discussions therein):
\be
   \Pi_{\rm FL} \sim \frac{1}{U}\! \left(\sinh^{-1} \frac{2\varepsilon_F^*}{v_F^* q} -\sinh^{-1} \frac{|\Omega|}{v_F^* q} \right). 
\ee
For typical momentum transfer $q\sim k_F$ and typical freqyency transfer $\Omega\sim T_{2e}$, we see that $\Pi_{\rm FL}\sim 1/U\ll \Pi_{\rm NFL}$, and thus
\begin{align}
    \Pi(\Omega)  \simeq   \frac{1}{2\sqrt{\pi}U} \, \ln \frac{U^2}{\varepsilon_F^2}.
\end{align}
In other words, the pairing of fermions in the heavy FL regime is actually mediated by the particle-particle fluctuations in the NFL regime. 

In the heavy FL regime, the eigenvalue problem for pairing is thus given by
\begin{align} \nonumber 
   & \lambda\Phi^*(k) = -\frac{\rho \mathcal{J} \Delta_{4e} Z^2}{\sqrt{\pi} U} \ln \left(\frac{U^2}{\varepsilon_F^2}\right) T k_F^{-d} \\&\times\sum_\Omega \int \frac{d\mathbf{q}}{(2\pi)^2} \,  \frac{\Phi(q)}{\Omega^2 + Z^2 \xi_\mathbf{q}^2},
\end{align}
where  $Z \sim 1/(k_F^{-d} \nu_0 U)$. The pairing temperature has the familiar BCS-form~\cite{Altland2010Condensed}:
\begin{align}
    T_{2e} \sim \frac{\varepsilon_F^2}{U} \exp\({-\frac{ U^2 }{|\rho|\mathcal{J}\Delta_{4e}\ln (U^2/\varepsilon_F^2)}}\#\). \label{Tc_Fl}
\end{align}
where again $\#$ stands for a nonuniversal $\mathcal{O}(1)$  number. 
In contrast with the conventional BCS theory, the phase transition is first-order.

\section{Discussion}
\label{sec:discussion}

In quest of the basic properties of a charge-$4e$ superconductor, in this work we constructed a mean-field toy model describing a Fermi surface  coupled to strong four-fermion interactions and charge-$4e$ superconducting order parameters. It is analytically tractable in the large-$N$ limit beyond perturbation theory.

We obtained several key results. First, in contrast with a charge-$2e$ superconductor, the Fermi surface remains gapless and hosts long-lived quasiparticles in its vicinity at low-energies. Despite that, unlike the Fermi liquid, the volume enclosed by the Fermi surface does not obey the Luttinger theorem. At higher-energies, we found that strong interaction effects make the system behave as a non-Fermi liquid. 

Second, in sharp contrast to a charge-$2e$ superconductor, {we found that the superfluid density for a charge-4e superconductor is in general smaller than the electron density, i.e., $n_s < n$, and is perturbatively small if $\Delta_{4e}$ is small compared to the strength of regular interactions.}

Third, we found that in the presence of correlation between $\mathcal{J}$ and $\Delta_{4e}$, the system admits a low-temperature pairing instability toward charge-$2e$ pairing phase via a first-order transition. The phase transition can either occur in the Fermi liquid regime or non-Fermi liquid regime, in the latter of which the pairing problem bears remarkable similarity to color superconductivity.

{Furthermore, we extended our analysis beyond the mean-field approach and demonstrate the validity of our results for an isolated charge-$4e$ superconductor that preserves its total charge.}

While the quantitative results inevitably depend on the microscopic model, we argued that the results above at a qualitative level are rather general for charge-$4e$ superconductors. Indeed, we showed that the smallness of the superfluid density and the low-temperature instability are directly connected with the gaplessness of the Fermi surface, which is an expected feature of a generic charge-$4e$ superconductor. 

We note that recently charge-$4e$ superconducting states have been proposed to exist in twisted bilayer graphene~\cite{Fernandes2021Charge4e}, the pair-density-wave state of cuprate superconudctors~\cite{Berg2009Charge}, and in the putative $\mathbb{Z}_4$ spin liquids~\cite{Barkeshli2013Gapless}. Also, experimental evidence of the charge-$4e$ superconductivity in the Kagome metal was reported in the recent preprint \cite{Ge2022Discovery}. It will be interesting to extend our results to these  contexts, either perturbatively or using numerical methods such as quantum Monte Carlo in the Majorana basis. 

\acknowledgments
We thank D.~Chowdhury, A.~V.~Chubukov, E.~Fradkin,  S.~Kivelson,  A.~I.~Pavlov, and an anonymous referee for illuminating discussions. N.G.~and Y.W.~are supported by startup funds at the University of Florida and by NSF under award number \addYW{DMR-2045781}. Part of this work was performed at the Aspen Center for Physics, which is supported by National Science Foundation grant PHY-1607611.

\begin{widetext}

\begin{appendix}

\section{Effective action and saddle-point analysis}\label{app:saddle-point}
In this Appendix we derive the saddle-point equations for the Hamiltonian (\ref{H}) in the main text. To simplify the presentation, we ignore the spatial dimensions of the problem since the procedure of the disorder averaging remains intact. The results of this Section are straightforward to generalize for the finite dimensional model.

We begin with a zero-dimensional version of the symmetrized interaction Hamiltonian (\ref{V}) in the main text
\begin{align} 
    H_{int} =&\sum_{\substack{i< j; k< l; i< k}}^N V_{ij;kl}, \label{app:H_0d} \\ 
    V_{ij;kl} =& \frac{ \mathcal{J}_{ij;kl}}{N^{3/2}}  \left(  \Psi^\dag_i i \sigma_y (\Psi_j^\dag)^T  \, \Psi_k^T i \sigma_y^T \Psi_l + h.c. \right)  +
    \frac{ {\Delta_{4e}}_{,ij;kl}}{N^{3/2}}  \left(  \Psi^\dag_i i \sigma_y (\Psi_j^\dag)^T \, \Psi^\dag_k i \sigma_y (\Psi_l^\dag)^T + h.c. \right) \\=& \frac{1}{N^{3/2}}  \( \mathcal{J}_{ij;kl} \left(  \varphi_{ij}^\dag  \, \varphi_{kl} + h.c. \right) + {\Delta_{4e}}_{,ij;kl} \left(  \varphi_{ij}^\dag  \, \varphi_{kl}^\dag + h.c. \right) \), 
\end{align}
where we introduce the bilinear operators $\varphi,\varphi^\dag$:
\begin{align}
\varphi_{ij}^\dag =& \Psi^\dag_i i \sigma_y (\Psi_j^\dag)^T = \psi^\dag_{i\uparrow} \psi^\dag_{j\downarrow} - \psi^\dag_{i\downarrow} \psi^\dag_{j\uparrow} = \varphi_{ji}^\dag, \label{app:phidag_ij} \\
\varphi_{kl} =& \Psi_k^T i \sigma_y^T \Psi_l = \psi_{k\downarrow} \psi_{l\uparrow}- \psi_{k\uparrow} \psi_{l\downarrow}=\varphi_{lk}. \label{app:phi_kl}
\end{align}
The coupling constants are given by the two sets $\lbrace J\rbrace$ and $\lbrace \Delta_{4e} \rbrace$ of real independent random Gaussian variables.
Both  $\lbrace J\rbrace$ and $\lbrace \Delta_{4e} \rbrace$ are drawn from the bivariate distribution (\ref{P_JK}) in the main text with a correlation parameter $-1<\rho<1$ between the two sets of random variables. 

We evaluate the disorder average of the partition function straightaway since the effects of the  replica symmetry breaking are negligible in the SYK model \cite{Wang2019Replica}. To do so, we integrate over each independent coupling coefficient with the bivariate Gaussian distribution (\ref{P_JK}):
\begin{align} \nonumber
    &\overline{z_{ij;kl}} \equiv \overline{\exp\bigg\lbrace -\int \!\! d\tau \, V_{ij;kl}(\tau)\bigg\rbrace} = \int  d \mathcal{J}_{ij;kl} \, d {\Delta_{4e}}_{,ij;kl}\, P(\mathcal{J}_{ij;kl},{\Delta_{4e}}_{ij;kl}) \\ \nonumber &\times \exp\bigg\lbrace \!-\mathcal{J}_{ij;kl} \int \!\! d\tau \(  \varphi_{ij}^\dag(\tau)   \varphi_{kl}(\tau) + h.c. \)
    - {\Delta_{4e}}_{,ij;kl} \int \!\! d\tau \(  \varphi_{ij}^\dag(\tau)   \varphi_{kl}^\dag(\tau) + h.c. \) \!\! \bigg\rbrace \\ \nonumber
    &= \exp\bigg\lbrace \frac{{\cal J}^2}{2N^3}\!\int \!\! d\tau d\tau' \!\( \varphi_{ij}^\dag(\tau) \varphi_{ji}(\tau')  \varphi_{lk}^\dag(\tau') \varphi_{kl}(\tau) + 
    \varphi_{ij}^\dag(\tau) \varphi_{ji}^\dag(\tau') \varphi_{kl}(\tau) \varphi_{lk}(\tau')  + h.c. \) \\ \nonumber 
    &+ \frac{\Delta_{4e}^2}{2N^3}\!\int \!\! d\tau d\tau' \!\(
    \varphi_{ij}^\dag(\tau) \varphi_{ji}^\dag(\tau') \varphi_{kl}^\dag(\tau) \varphi_{lk}^\dag(\tau') + \varphi_{ij}^\dag(\tau) \varphi_{ji}(\tau') \varphi_{kl}^\dag(\tau) \varphi_{lk}(\tau') +h.c. \) \\ 
    &+ \rho \frac{{\cal J}\Delta_{4e}}{N^3}\!\int \!\! d\tau d\tau' \! \(
    \varphi_{ij}^\dag(\tau) \varphi_{ji}^\dag(\tau') \varphi_{kl}(\tau) \varphi_{lk}^\dag(\tau') + \varphi_{ij}^\dag(\tau) \varphi_{ji}(\tau') \varphi_{kl}(\tau)\varphi_{lk}(\tau') + h.c. \) \!\!
    \bigg\rbrace, \label{app:dav}
\end{align}
where $\overline{(\dots)}$ denotes the disorder average. The interacting part of the effective action can be found from 
\begin{align}
    S_{int} = - \sum_{\substack{i< j; k< l; i< k}}^N V_{ij;kl} \ln \overline{z_{ij;kl}} \simeq -\frac{1}{8} \sum_{i,j,k,l=1}^N \ln \overline{z_{ij;kl}}, \label{app:S_int_def}
\end{align}
where the $\bar{z}$ terms with the coinciding indices are neglected in the large-$N$ limit.

To decouple the $8$-fermion interactions in Eqs.~(\ref{app:dav},\ref{app:S_int_def}),
we introduce the self-energies and Green's functions 
via the following identities: 
\begin{align} 
    \mathbb{1}= & \int \! \mathcal{D}\Sigma\mathcal{D}G \exp\left\lbrace \sum_{\sigma=\uparrow\downarrow} \int  \!\! d \tau  d \tau' \, \Sigma_{\sigma\sigma}(\tau,\tau')\left(N G_{\sigma\sigma}(\tau'\!,\tau) - \sum_{i=1}^N \bar{\psi}_{i\sigma}(\tau)\psi_{i\sigma}(\tau') \right) \right\rbrace, \label{app:G_def}\\ 
    \mathbb{1}= & \int \! \mathcal{D}\Phi\mathcal{D}F^+ \exp\left\lbrace  \int \!\! d \tau  d \tau' \, \Phi_{\uparrow\downarrow}(\tau,\tau')\left(N F^+_{\downarrow\uparrow}(\tau'\!,\tau) - \sum_{i=1}^N \bar{\psi}_{i\uparrow}(\tau)\bar{\psi}_{i\downarrow}(\tau') \right) \right\rbrace, \label{app:F+_def}\\ 
    \mathbb{1}= & \int \! \mathcal{D}\Phi^+ \mathcal{D}F \exp\left\lbrace  \int \!\! d \tau d \tau' \, \Phi^+_{\downarrow\uparrow}(\tau,\tau')\left(N F_{\uparrow\downarrow}(\tau'\!,\tau) - \sum_{i=1}^N \psi_{i\downarrow}(\tau)\psi_{i\uparrow}(\tau') \right) \right\rbrace,
    \label{app:F_def}
\end{align}  
where we assume $G_{\uparrow\downarrow}=G_{\downarrow\uparrow}=0$ and $F_{\uparrow\uparrow}=F_{\downarrow\downarrow} = F_{\uparrow\uparrow}^+=F_{\downarrow\downarrow}^+ =0$. The anomalous contributions (\ref{app:F+_def},\ref{app:F_def}) to the effective action are introduced in accordance with definition of the anomalous blocks of the Gor'kov's Greens's function (\ref{F_def},\ref{F+_def}) in the main text. Using the identities (\ref{app:G_def}-\ref{app:F_def}), we compute the effective action (\ref{app:S_int_def}).
Indeed, the contributions to the effective action after the disorder average (\ref{app:dav}) can be expressed in terms of $4$-fermion products:
\begin{align} \nonumber
    &\sum_{i,j=1}^N\varphi_{ij}^\dag(\tau) \varphi_{ji}(\tau') = \sum_{i,j=1}^N \Psi^\dag_i(\tau) i \sigma_y (\Psi_j^\dag(\tau))^T \Psi_j^T(\tau') i \sigma_y^T \Psi_i(\tau') \\&= N\sum_{i=1}^N \begin{pmatrix}\bar{\psi}_{i\uparrow} & \bar{\psi}_{i\downarrow} \end{pmatrix}_\tau  \begin{pmatrix} G_{\downarrow \downarrow}(\tau'\!,\tau) & 0\\ 0 & G_{\uparrow \uparrow}(\tau'\!,\tau) \end{pmatrix}  \begin{pmatrix} \psi_{i\uparrow} \\ \psi_{i\downarrow} \end{pmatrix}_{\tau'} = 2 N^2 G_{\uparrow \uparrow}(\tau'\!,\tau) G_{\downarrow \downarrow}(\tau'\!,\tau),  \label{app:phidagphi}\\  \nonumber
    &\sum_{i,j=1}^N\varphi_{ij}^\dag(\tau) \varphi_{ji}^\dag(\tau') = \sum_{i,j=1}^N \Psi^\dag_i(\tau) i \sigma_y (\Psi_j^\dag(\tau))^T \Psi^\dag_j(\tau') i \sigma_y (\Psi_i^\dag(\tau'))^T \\&= N\sum_{i=1}^N \begin{pmatrix}\bar{\psi}_{i\uparrow} & \bar{\psi}_{i\downarrow} \end{pmatrix}_\tau  \begin{pmatrix} 0 & -F_{\downarrow\uparrow}^+(\tau,\tau')\\ F_{\downarrow\uparrow}^+(\tau'\!,\tau) & 0 \end{pmatrix}  \begin{pmatrix} \bar{\psi}_{i\uparrow} \\ \bar{\psi}_{i\downarrow} \end{pmatrix}_{\tau'} = - 2N^2 F^+_{\downarrow\uparrow}(\tau'\!,\tau) F^+_{\downarrow\uparrow}(\tau,\tau'),  \\ \nonumber
    &\sum_{i,j=1}^N\varphi_{ij}(\tau) \varphi_{ji}(\tau') = \sum_{i,j=1}^N \Psi_i^T(\tau) i \sigma_y^T \Psi_j(\tau) \Psi_j^T(\tau') i \sigma_y^T \Psi_i(\tau') \\&= N\sum_{i=1}^N \begin{pmatrix}\psi_{i\uparrow} & \psi_{i\downarrow} \end{pmatrix}_\tau \begin{pmatrix} 0 & F_{\uparrow\downarrow}(\tau'\!,\tau)\\ -F_{\uparrow\downarrow}(\tau,\tau') & 0 \end{pmatrix}  \begin{pmatrix} \psi_{i\uparrow} \\ \psi_{i\downarrow} \end{pmatrix}_{\tau'} = - 2N^2 F_{\uparrow\downarrow}(\tau'\!,\tau) F_{\uparrow\downarrow}(\tau,\tau'). \label{app:phiphi}
\end{align}

Substituting Eqs.~(\ref{app:phidagphi}-\ref{app:phiphi}) into Eqs.~(\ref{app:dav},\ref{app:S_int_def}), we derive the effective action for the Hamiltonian (\ref{app:H_0d}):
\begin{align} \nonumber
    S =& \sum_{i=1}^N \int d\tau d\tau' \begin{pmatrix}\bar{\psi}_{i\uparrow} & \psi_{i\downarrow}\end{pmatrix}_\tau  \begin{pmatrix}\delta_{\tau\tau'}\partial_\tau + \Sigma_{\uparrow\uparrow}(\tau,\tau') & \Phi_{\uparrow\downarrow}(\tau,\tau') \\ \Phi^+_{\downarrow\uparrow}(\tau,\tau') & \delta_{\tau\tau'}\partial_\tau - \Sigma_{\downarrow\downarrow}(\tau'\!,\tau) \end{pmatrix} \begin{pmatrix}\psi_{i\uparrow} \\ \bar{\psi}_{i\downarrow}\end{pmatrix}_{\tau'} \\ \nonumber
    - & N \int d\tau d\tau' \bigg( \sum_{\sigma=\uparrow\downarrow}\Sigma_{\sigma\sigma}(\tau,\tau') G_{\sigma\sigma}(\tau'\!,\tau) + \Phi_{\uparrow\downarrow}(\tau,\tau') F^+_{\downarrow\uparrow}(\tau'\!,\tau) + \Phi^+_{\downarrow\uparrow}(\tau,\tau') F_{\uparrow\downarrow}(\tau'\!,\tau) \\ \nonumber 
    +& \frac{{\cal J}^2}{2} \big(G_{\uparrow\uparrow}(\tau,\tau')G_{\uparrow\uparrow}(\tau'\!,\tau)G_{\downarrow\downarrow}(\tau,\tau')G_{\downarrow\downarrow}(\tau'\!,\tau) + F^+_{\downarrow\uparrow}(\tau,\tau')F^+_{\downarrow\uparrow}(\tau'\!,\tau) F_{\uparrow\downarrow}(\tau,\tau')F_{\uparrow\downarrow}(\tau'\!,\tau) \big) \\ \nonumber
    +&\frac{\Delta_{4e}^2}{4} \big( G_{\uparrow\uparrow}(\tau,\tau')^2 G_{\downarrow\downarrow}(\tau,\tau')^2 + G_{\uparrow\uparrow}(\tau'\!,\tau)^2 G_{\downarrow\downarrow}(\tau'\!,\tau)^2 + F^+_{\downarrow\uparrow}(\tau,\tau')^2 F^+_{\downarrow\uparrow}(\tau'\!,\tau)^2  \\ \nonumber +& F_{\uparrow\downarrow}(\tau,\tau')^2F_{\uparrow\downarrow}(\tau'\!,\tau)^2 \big) - \rho\, \frac{{\cal J}\Delta_{4e}}{2} \big( G_{\uparrow\uparrow}(\tau,\tau') G_{\downarrow\downarrow}(\tau,\tau') + G_{\uparrow\uparrow}(\tau'\!,\tau) G_{\downarrow\downarrow}(\tau'\!,\tau) \big) \\ \times & \big( F^+_{\downarrow\uparrow}(\tau,\tau')F^+_{\downarrow\uparrow}(\tau'\!,\tau) + F_{\uparrow\downarrow}(\tau,\tau')F_{\uparrow\downarrow}(\tau'\!,\tau) \big) \bigg).
\end{align}
Integrating over fermions we get 
\begin{align} \nonumber
    S =& -N \sum_\omega \ln \bigg( \left(i\omega -\Sigma_{\uparrow\uparrow}(\omega) \right)\left(i\omega +\Sigma_{\downarrow\downarrow}(-\omega) \right) -  \Phi_{\uparrow\downarrow}(\omega)\Phi^+_{\downarrow\uparrow}(\omega)  \bigg) \\  \nonumber
    - & N \int d\tau d\tau' \bigg( \sum_{\sigma=\uparrow\downarrow}\Sigma_{\sigma\sigma}(\tau,\tau') G_{\sigma\sigma}(\tau'\!,\tau) + \Phi_{\uparrow\downarrow}(\tau,\tau') F^+_{\downarrow\uparrow}(\tau'\!,\tau) + \Phi^+_{\downarrow\uparrow}(\tau,\tau') F_{\uparrow\downarrow}(\tau'\!,\tau) \\ \nonumber 
    +& \frac{{\cal J}^2}{2} \big(G_{\uparrow\uparrow}(\tau,\tau')G_{\uparrow\uparrow}(\tau'\!,\tau)G_{\downarrow\downarrow}(\tau,\tau')G_{\downarrow\downarrow}(\tau'\!,\tau) + F^+_{\downarrow\uparrow}(\tau,\tau')F^+_{\downarrow\uparrow}(\tau'\!,\tau) F_{\uparrow\downarrow}(\tau,\tau')F_{\uparrow\downarrow}(\tau'\!,\tau) \big) \\ \nonumber
    +&\frac{\Delta_{4e}^2}{4} \big( G_{\uparrow\uparrow}(\tau,\tau')^2 G_{\downarrow\downarrow}(\tau,\tau')^2 + G_{\uparrow\uparrow}(\tau'\!,\tau)^2 G_{\downarrow\downarrow}(\tau'\!,\tau)^2 + F^+_{\downarrow\uparrow}(\tau,\tau')^2 F^+_{\downarrow\uparrow}(\tau'\!,\tau)^2  \\ \nonumber +& F_{\uparrow\downarrow}(\tau,\tau')^2F_{\uparrow\downarrow}(\tau'\!,\tau)^2 \big) - \rho\, \frac{{\cal J}\Delta_{4e}}{2} \big( G_{\uparrow\uparrow}(\tau,\tau') G_{\downarrow\downarrow}(\tau,\tau') + G_{\uparrow\uparrow}(\tau'\!,\tau) G_{\downarrow\downarrow}(\tau'\!,\tau) \big) \\ \times & \big( F^+_{\downarrow\uparrow}(\tau,\tau')F^+_{\downarrow\uparrow}(\tau'\!,\tau) + F_{\uparrow\downarrow}(\tau,\tau')F_{\uparrow\downarrow}(\tau'\!,\tau) \big) \bigg), \label{app:S_eff}
\end{align}
where $\sum\limits_\omega$ denotes the summation over the Matsubara frequencies.

The effective action \eqref{app:S_eff} has a saddle-point in the large-$N$ limit, therefore, we take the variation of the effective action with respect to the bi-local fields $G$, $F$, and $F^*$ that leads to the first set of the Schwinger-Dyson equations:  
\begin{align} \nonumber
    \frac{\delta S}{\delta G_{\uparrow\uparrow}} = 0 \quad \Rightarrow \quad \Sigma_{\uparrow\uparrow}(\tau) =& - \mathcal{J}^2 G_{\uparrow\uparrow}(\tau) G_{\downarrow\downarrow}(\tau) G_{\downarrow\downarrow}(-\tau) - \Delta_{4e}^2 G_{\downarrow\downarrow}(-\tau)^2 G_{\uparrow\uparrow}(-\tau) \\ 
    &+ \rho\, \mathcal{J} \Delta_{4e} \(F^+_{\downarrow\uparrow}(\tau)F^+_{\downarrow\uparrow}(-\tau) + F_{\uparrow\downarrow}(\tau)F_{\uparrow\downarrow}(-\tau)\) G_{\downarrow\downarrow}(-\tau), \\  \nonumber
    \frac{\delta S}{\delta G_{\downarrow\downarrow}} = 0 \quad \Rightarrow \quad \Sigma_{\downarrow\downarrow}(\tau) =& - \mathcal{J}^2 G_{\downarrow\downarrow}(\tau) G_{\uparrow\uparrow}(\tau) G_{\uparrow\uparrow}(-\tau) - \Delta_{4e}^2 G_{\uparrow\uparrow}(-\tau)^2 G_{\downarrow\downarrow}(-\tau) \\ 
    &+ \rho\, \mathcal{J} \Delta_{4e} \(F^+_{\downarrow\uparrow}(\tau)F^+_{\downarrow\uparrow}(-\tau) + F_{\uparrow\downarrow}(\tau)F_{\uparrow\downarrow}(-\tau)\) G_{\uparrow\uparrow}(-\tau), \\ \nonumber
    \frac{\delta S}{\delta F^+_{\downarrow\uparrow}} = 0  \quad \Rightarrow \quad \Phi_{\uparrow\downarrow}(\tau) = & - \mathcal{J}^2 F_{\downarrow\uparrow}^+(\tau) F_{\uparrow\downarrow}(\tau) F_{\uparrow\downarrow}(-\tau) - \Delta_{4e}^2 F_{\downarrow\uparrow}^+(\tau)^2 F_{\downarrow\uparrow}^+(-\tau) \\ 
    &+ \rho\, \mathcal{J} \Delta_{4e} \( G_{\uparrow\uparrow}(\tau) G_{\downarrow\downarrow}(\tau) + G_{\uparrow\uparrow}(-\tau) G_{\downarrow\downarrow}(-\tau) \) F_{\downarrow\uparrow}^+(\tau), \\ \nonumber
    \frac{\delta S}{\delta F_{\uparrow\downarrow}} = 0  \quad \Rightarrow \quad \Phi^+_{\downarrow\uparrow}(\tau) = & - \mathcal{J}^2 F_{\uparrow\downarrow}(\tau) F_{\downarrow\uparrow}^+(\tau) F_{\downarrow\uparrow}^+(-\tau) - \Delta_{4e}^2 F_{\uparrow\downarrow}(\tau)^2 F_{\uparrow\downarrow}(-\tau) \\ 
    &+ \rho\, \mathcal{J} \Delta_{4e} \( G_{\uparrow\uparrow}(\tau) G_{\downarrow\downarrow}(\tau) + G_{\uparrow\uparrow}(-\tau) G_{\downarrow\downarrow}(-\tau) \) F_{\uparrow\downarrow}(\tau),
\end{align}
where we have used translational invariance in the arguments of the bilocal fields $\tau,\tau' \to \tau - \tau'\to \tau$. 
The second set of the Schwinger-Dyson equations is 
\begin{align}
    \frac{\delta S}{\delta \Sigma_{\uparrow\uparrow}} =& 0 \quad \Rightarrow \quad  G_{\uparrow\uparrow}(\omega) = \frac{i\omega+\Sigma_{\downarrow\downarrow}(-\omega)}{\left(i\omega-\Sigma_{\uparrow\uparrow}(\omega)\right)\left(i\omega+\Sigma_{\downarrow\downarrow}(-\omega)\right)-\Phi_{\uparrow\downarrow}(\omega) \Phi_{\downarrow\uparrow}^+(\omega)}, \\
    \frac{\delta S}{\delta \Sigma_{\downarrow\downarrow}} =& 0 \quad \Rightarrow \quad  G_{\downarrow\downarrow}(-\omega) = \frac{-i\omega+\Sigma_{\uparrow\uparrow}(\omega)}{\left(i\omega-\Sigma_{\uparrow\uparrow}(\omega)\right)\left(i\omega+\Sigma_{\downarrow\downarrow}(-\omega)\right)-\Phi_{\uparrow\downarrow}(\omega) \Phi_{\downarrow\uparrow}^+(\omega)}, \\
    \frac{\delta S}{\delta F_{\uparrow\downarrow}} =& 0 \quad \Rightarrow \quad  F_{\downarrow\uparrow}^+(\omega) = \frac{\Phi_{\downarrow\uparrow}^+(\omega)}{\left(i\omega-\Sigma_{\uparrow\uparrow}(\omega)\right)\left(i\omega+\Sigma_{\downarrow\downarrow}(-\omega)\right)-\Phi_{\uparrow\downarrow}(\omega) \Phi_{\downarrow\uparrow}^+(\omega)}, \\
    \frac{\delta S}{\delta F^+_{\downarrow\uparrow}} =& 0 \quad \Rightarrow \quad  F_{\uparrow\downarrow}(\omega) = \frac{\Phi_{\uparrow\downarrow}(\omega)}{\left(i\omega-\Sigma_{\uparrow\uparrow}(\omega)\right)\left(i\omega+\Sigma_{\downarrow\downarrow}(-\omega)\right)-\Phi_{\uparrow\downarrow}(\omega) \Phi_{\downarrow\uparrow}^+(\omega)}.
\end{align}

For the self-energies we imply $\Sigma_{\uparrow \uparrow}=\Sigma_{\downarrow \downarrow} =\Sigma$ under spin-rotation symmetry as stated in Eq.~\eqref{G_def} in the main text. Here $\Phi_{\uparrow\downarrow}$ and $\Phi_{\downarrow\uparrow}^+$ are the pairing vertexes in the 
spin-singlet channel. Assuming that the energetically favorable pairing vertex 
is even in frequency, we have  $\Phi_{\uparrow\downarrow} = -\Phi_{\downarrow\uparrow} = \Phi$ and $\Phi^+_{\downarrow\uparrow} = -\Phi^+_{\uparrow\downarrow} = \Phi^*$.
The same relations hold for the Green's functions which are introduced in Eqs.~(\ref{F_def},\ref{F+_def}) in the main text. As such, we simplify the Schwinger-Dyson equations to
\begin{align} 
    \Sigma(\tau) =& - \mathcal{J}^2 G(\tau)^2 G(-\tau) - \Delta_{4e}^2 G(-\tau)^3 +  \rho \,\mathcal{J}\Delta_{4e}  \left( F^*(\tau)^2 + F(\tau)^2 \right) G(-\tau), \\ 
    \Phi^*(\tau) =& - \mathcal{J}^2 F^*(\tau)^2 F(\tau) - \Delta_{4e}^2 F(\tau)^3 + \rho \,\mathcal{J}\Delta_{4e}  \left( G(\tau)^2 + G(-\tau)^2 \right) F(\tau)
\end{align}
and 
\begin{align}
    G(\omega) =& \frac{i\omega+\Sigma(-\omega)}{\left(i\omega-\Sigma(\omega)\right)\left(i\omega+\Sigma(-\omega)\right)-\Phi^*(\omega)\Phi(\omega)}, \\
    F^*(\omega) =& \frac{\Phi^*(\omega)}{\left(i\omega-\Sigma(\omega)\right)\left(i\omega+\Sigma(-\omega)\right)-\Phi^*(\omega)\Phi(\omega)}.
\end{align}
It is straightforward to generalize these saddle point equations to Eqs.~(\ref{Sigma}-\ref{F}) in the main text for the finite-dimensional model. One can also derive the Schwinger-Dyson equations by variation of the effective action (\ref{S}) in the main text, where the pairing and spin symmetries of the considered saddle-point have been already accounted.


\section{Derivation of the normal-state Green's function in 2d} \label{app:IR}

In this Appendix, we derive the normal-state self-energy and Green's function. We self-consistently show that there is an emergent energy scale $\varepsilon_F^*=\varepsilon_F^2/U$. At frequencies and temperatures below this scale, i.e., $\omega, T \ll \varepsilon_F^*$ , the system is a heavy Fermi liquid, and for $ \varepsilon_F^*\ll \omega, T \ll U$, the system behaves as a non-Fermi liquid.

For simplicity of presentation, our derivation below is performed for the 2d case. It is however straightforward to generalize our derivation to a $d>2$ case, by including additional angular directions for the momentum integral.


\subsection{Heavy Fermi liquid at $T\ll \varepsilon_F^2/U$}
\label{app:B1}

At low temperatures, we take the ansatz that the low-energy form of the  Green's function takes the form of \eqref{G_4e} in the main text:
\begin{align}
    G(\omega, \mathbf{k}) = \frac{Z}{i\omega -\xi^*_{\mathbf{k}}}, \quad \omega, T \ll \varepsilon^*_F. \label{G_QP_anz}
\end{align}
Here $Z$ is the quasiparticle residue, $\xi^*_{\mathbf{k}}\approx \mathbf{v}_F^* \cdot \mathbf{k}$ is the renormalized dispersion, $v_F^*$ is the renormalized Fermi velocity, and $\varepsilon^*_F$ is the renormalized Fermi energy. 

We begin with evaluation of the particle-particle bubble (\ref{PP_4e}) with the ansatz (\ref{G_QP_anz}):
\begin{align} 
    \Pi(\Omega, \mathbf{q}) =& \Pi_0 + \widetilde{\Pi}(\Omega, \mathbf{q})
\nonumber \\
=& \Pi_0 +Z^2 k_F^{-2}\int_{-\varepsilon^*_{F}}^{\varepsilon^*_{F}} \frac{d\omega}{2\pi} \int \!\! \frac{d\mathbf{k}}{(2\pi)^2} \frac{1}{\frac{i \Omega}{2} + i\omega -\xi^*_{\frac{\mathbf{q}}{2} + \mathbf{k}}}   \frac{1}{\frac{i \Omega}{2} - i\omega -\xi^*_{\frac{\mathbf{q}}{2} - \mathbf{k}}}.
\label{eq:bubble}
\end{align}
The first term $\Pi_0$ here comes from fermions with higher energies, e.g., in the NFL regime. For our purposes here the relevant $\Omega, v_F^*q \ll \varepsilon_F^2/U$, and thus these high-energy contributions are constant in $\Omega, v_F^*q$.

We consider low-lying excitations and, hence, expand the dispersion $\xi^*_{\frac{\mathbf{q}}{2} \pm \mathbf{k}} \simeq \xi^*_\mathbf{k} \pm \frac{\mathbf{v}_F^*\cdot \mathbf{q}}{2} $ in the excitation momentum $\mathbf{q}$. Furthermore, we introduce the density of states $\nu^*$, which is constant $\nu^* = k_F/(\pi v_F^*)$ in two dimensions, and replace the integral over fermion momentum $\mathbf{k}$ with the energy $\varepsilon$ integral:
\begin{align} \nonumber
    \widetilde{\Pi}(\Omega, \mathbf{q}) \simeq & \nu^* Z^2 k_F^{-2}\!\! \int_{-\varepsilon^*_{F}}^{\varepsilon^*_{F}} \frac{d\omega}{2\pi} \int_0^{2\pi} \! \frac{d\theta}{2\pi}  \int_{-\infty}^{+\infty} \!\! d \varepsilon  \frac{1}{\frac{i\Omega}{2} + i\omega -\varepsilon -\frac{v_F^*q}{2}\cos\theta}\frac{1}{\frac{i\Omega}{2} - i\omega -\varepsilon +\frac{v_F^*q}{2}\cos\theta} \\ \nonumber
    = & \nu^* Z^2 k_F^{-2}\!\! \int_{\frac{|\Omega|}{2}}^{\varepsilon^*_{F}} \!\! d\omega \int_0^{2\pi} \!\! \frac{d\theta}{2\pi} \frac{4 \omega}{4\omega^2 + {v_F^*}^2 q^2 \cos^2\theta} = 2 \nu^* Z^2 k_F^{-2}\int_{\frac{|\Omega|}{2}}^{\varepsilon^*_{F}}     \frac{d\omega}{\sqrt{4\omega^2 + {v_F^*}^2 q^2}} \\  
    =& \nu^* Z^2 k_F^{-2}\! \left(\sinh^{-1} \frac{2\varepsilon_F^*}{v_F^* q} -\sinh^{-1} \frac{|\Omega|}{v_F^* q} \right). \label{PP}
\end{align}
Here $\theta$ is the angle between $\mathbf{q}$ and $\mathbf{k}${, $q=|{\bf q}|$,} and
$v_F^*$ is the renormalized Fermi velocity.  We close the contour in the upper half complex plane when evaluating the integral over energy $\varepsilon$. For the frequencies $-|\Omega|/2 < \omega < |\Omega|/2$ there are two contributing poles that cancel each other, while both $-\varepsilon_F^* \leq \omega \leq -|\Omega|/2$ and $|\Omega|/2 \leq \omega \leq \varepsilon_F^*$ regions contribute one pole each. 

The particle-particle bubble (\ref{PP}) is an even function of frequency and momenta $\widetilde{\Pi}(\Omega, \mathbf{q})=\widetilde{\Pi}(-\Omega, -\mathbf{q})$. Furthermore, the constant piece $\Pi_0$ leads to a constant real piece after the integral, contributing to the renormaliztion of the chemical potential. 
Therefore, for the frequency and momentum dependence of the self-energy $\Sigma$ we only need $\widetilde{\Pi}(\Omega,\mathbf{q})$ in the particle-particle bubble. From Eq. (\ref{Sigma_4e}) we get
\begin{align} 
    \Sigma(\omega,\mathbf{k})  = & - Z U^2  k_F^{-2}\!\! \int_{-\varepsilon_F^*}^{\varepsilon_F^*}  \frac{d\Omega}{2\pi} \int  \frac{d\mathbf{q}}{(2\pi)^2} \,\, \frac{\widetilde{\Pi}(\Omega, \mathbf{q})}{i \Omega - i\omega -\xi^*_{\mathbf{q}-\mathbf{k}-\mathbf{k}_F}}, \label{Sigma_IR}
\end{align}
where $\mathbf{k}$ is the deviation from the Fermi surface. 

In order to self-consistently solve for $Z$ and $v_F^*$, we separately consider the frequency and momentum dependence of the self-energy $\Sigma$ at leading order. For the frequency dependence, we get 
\be
\Sigma(\omega,\mathbf{k}=0) = -iU^2Z^3\nu^*k_F^{- 4}\!\int\frac{qdq\, d\Omega}{4\pi^2} \frac{\sgn(\omega-\Omega)}{\sqrt{(\omega-\Omega)^2+v^{*2}q^2}}\(\sinh^{-1}\frac{2 \varepsilon_F^*}{v_F^*q}-\sinh^{-1}\frac{|\Omega|}{v_F^*q}\).
\ee
The leading, linear in $\omega$, contribution comes from small internal frequency $\Omega\ll v_F^* q$:
\be
\Sigma(\omega, \mathbf{k}=0) \approx -i\omega \times \frac{2U^2Z^3\nu^*}{v_F^*k_F^{{ 4}}} \int \frac{dq}{4\pi^2} \sinh^{-1}\frac{{2} \varepsilon_F^*}{v_F^* q}\sim i\omega \times \frac{U^2Z^3\nu^*}{v_F^* k_F^{{3}}}, \label{app:eq_sigma_omega}
\ee
while the opposite limit $v_F^*q\ll \Omega$ yields a frequency dependence $\omega^2\ln\omega$ \cite{Chowdhury2018Translationally}.

In the strong coupling limit, the self-energy effects dominate the fermion Green's function:
\be
Z^{-1} \sim \frac{U^2Z^3\nu^* }{v_F^*k_F^3}.
\ee
As we shall see later, the renormalized Fermi velocity and density of states are given by $v_F^*\sim Z v_F$ and $\nu^*\sim {k_F^2/(Z \varepsilon_F)} \sim k_F/(Z v_F)$, which leads to 
\be
Z \sim \frac{\varepsilon_F}{U} {\, \ll 1}, 
\label{eq:40}
\ee
{where we have used $\varepsilon_F \sim v_F k_F$.}

Let us now evaluate the $k$-dependence of $\Sigma$ near the Fermi surface:
\be
\Sigma(\omega=0,\mathbf{k}) = U^2 Z^3 \nu^*k_F^{- 4}\!\int\!\frac{d{\bf q}\, d\Omega}{8\pi^3} \frac{v_F^*(q_x-k_x)}{{\Omega^2+v_F^{*2}(q_x-k_x)^2}}\!\(\sinh^{-1}\frac{ {2} \varepsilon_F^*}{v_F^*q}-\sinh^{-1}\frac{|\Omega|}{v_F^*q}\)\!,
\ee
where $q_x ,k_x$ refer to the momentum components in the direction perpendicular to the local patch of the Fermi surface. This is justified for $q_y^2\ll {k_F}(q_x-k_x)$, which is the dominant regime that contributes to the integral.

The first term in the parenthesis, after $\Omega$ integration becomes $\propto \sgn(q_x-k_x)$. Aside from a constant term, to leading order in $k$ (defined as the deviation from the Fermi surface), it is 
\be
\sim  \int_{-k}^{k} dq_x \int^{\sqrt{k_F k}}_{0} dq_y \ln\frac{{4} \varepsilon_F^*}{v_F^* q},
\ee
By a simple dimensional analysis, this contribution to $\Sigma$ comes as a higher-order term  $\sim \sgn(k_x) |k_x|^{3/2}\ln{|k_x|}$.

The frequency integral with the second term in the parenthesis yieds a log, cut by $v_F^*q$ above and $v_F|q_x-k_x|$ below. We have
\be
\Sigma(\omega=0,k_x)\approx { {\rm const} -} U^2Z^3\nu^*k_F^{-{4}} \int\frac{d{\bf q}}{{ 4}\pi^3}\frac{q_x-k_x}{q}\ln\frac{q}{|q_x-k_x|},
\label{eq:44}
\ee
which agrees with Eq.\ (D14) of Ref.~\cite{Chowdhury2018Translationally} using a slightly different method. However, as we show below the evaluation of the integral in Eq.~\eqref{eq:44} leads to a result with an opposite sign. Interestingly, the correct sign is crucial for our purposes as it ensures the superfluid density is positive and consistent with the $f$-sum rule \cite{Mahan2000Many}.

The leading contribution to Eq.~\eqref{eq:44} comes from the region in which $q_y\gg q_x$ such that the logarithm in the integrand is large, and in which $|q_x-k_x|\gg q_y^2/k_F$  such that the dispersion is linear. Keeping terms up to linear-in-$k$  order
\begin{align}
\Sigma(\omega=0,k_x)\approx&   {\rm const} - k_x\times \frac{U^2Z^3\nu^*}{\pi^3k_F^{{ 4}}}\int_{0}^{k_F} \frac{dq_y}{q_y} \int^{q_y}_{q_y^2/k_F}{dq_x}\[\ln\frac{q_x}{q_y}+1\]\nonumber\\
=& {\rm const} + k_x\times \frac{U^2 Z^3\nu^*}{\pi^3 k_F^{4}}\int_{0}^{k_F} {dq_y} \, { \frac{q_y}{k_F} }\ln\frac{k_F}{q_y}.
\end{align}
While the last integral is only accurate up to a non-universal factor, it is clear that it is positive, contrary to the result in Ref.~\cite{Chowdhury2018Translationally}.

We thus obtain
\be
\Sigma(\omega=0,k_x) \approx  { {\rm const} +}  {\beta} \, k_x \frac{ U^2Z^3\nu^*}{k_F^{{3}}} \equiv \textrm{const} + {\beta} \, 
v_F k_x, \label{app:eq_sigma_kx}
\ee
{where we use Eq.~\eqref{eq:40} and the fact that $\nu^* = \nu_0/Z$. The non-universal coefficient $0<\beta=\mathcal{O}(1)$ may depend on the geometry of the specific Fermi surface.}
The result that $\beta=\mathcal{O}(1)$ also confirms the 
{expression for the renormalized Fermi velocity} $v_F^* = Z v_F$ we previously used in deriving $Z$.

From the above results on $Z$ and $\beta$, we see that the renormalized Fermi energy is given by
\be
\varepsilon_F^* = Z \varepsilon_F \sim \frac{\varepsilon_F^2}{U}.
\ee
Similar to the standard case, the Fermi-liquid solution is only valid if 
\be
T\ll {\varepsilon_F^*} \sim \frac{\varepsilon_F^2}{U}.
\ee

To confirm the analytical estimates for the quasiparticle renormalization, we use the result (\ref{PP}) and numerically evaluate the self-energy (\ref{Sigma_IR}) for a 2d circular Fermi surface for arbitrary excitation momentum $\mathbf{q}$. 
Computing numerically the frequency and momentum integrals in Eq. (\ref{Sigma_IR}) for electrons with the quadratic dispersion {and Fermi energy $\varepsilon_F = k_F^2/(2m)$}, we derive the self-energy in the leading order in frequency and momentum:
\begin{align}
    \Sigma(\omega, k_x)  \simeq {\rm const} - i \omega \, \frac{\alpha_1}{(2\pi)^2}  \, Z  k_F^{-4} (\nu_0 U)^2 + \frac{v_F}{2}  k_x \, \frac{\alpha_2}{(2\pi)^2} \, Z^2  k_F^{-4} (\nu_0 U)^2, \label{Sigma_IR_lead}
\end{align}
where we use that $\nu^* = k_F^2/(2 \pi \varepsilon_F^*)$ and express it in terms of the density of states of the free fermions $\nu^* = \nu_0/Z$. Here we have chosen the Fermi momemtum $\mathbf{k}_F$ aligned with the $k_x$ component of the electron's momentum $\mathbf{k} = \lbrace k_x, k_y \rbrace$.
The numerical coefficients $\alpha_1 \approx 13.1$ and $\alpha_2 \approx 4$ are found from the frequency and momentum dependence of the self-energy (\ref{Sigma_IR}) shown in Fig. \ref{fig:Sigma_FL} in the low frequency ($\omega \leq 0.1 \varepsilon^*_F$ with $11$ points) and small momentum ($k_x \leq 0.05 k_F$ with $11$ points) 
Comparison of the imaginary part of the self-energy (\ref{Sigma_IR_lead}) in a strong coupling limit ($k_F^{-2} \nu_0 U \gg 1$) with the quasiparticle ansatz (\ref{G_QP_anz}) fixes the quasiparticle residue $Z$:
\begin{align}
    Z = \frac{2\pi}{\sqrt{\alpha_1}} \, \frac{1}{k_F^{-2} \nu_0 U} \approx 10.9 \, \frac{\varepsilon_F}{U}. \label{app:Z_num}
\end{align}
Substituting the quasiparticle residue \eqref{app:Z_num} into Eq.~\eqref{Sigma_IR_lead}, we obtain 
\begin{align}
    \Sigma(\omega, k_x)  \simeq {\rm const} - i \, Z^{-1} \omega  + \frac{\alpha_2}{2\alpha_1} \, v_F k_x,
\end{align}
which gives us the self-energy expression \eqref{eq:beta1} in the main text with 
$$\beta = 2\alpha_2/(\alpha_1) \approx 0.6.$$ 

\begin{figure}[t!!]
\center
\includegraphics[width=0.464\linewidth]{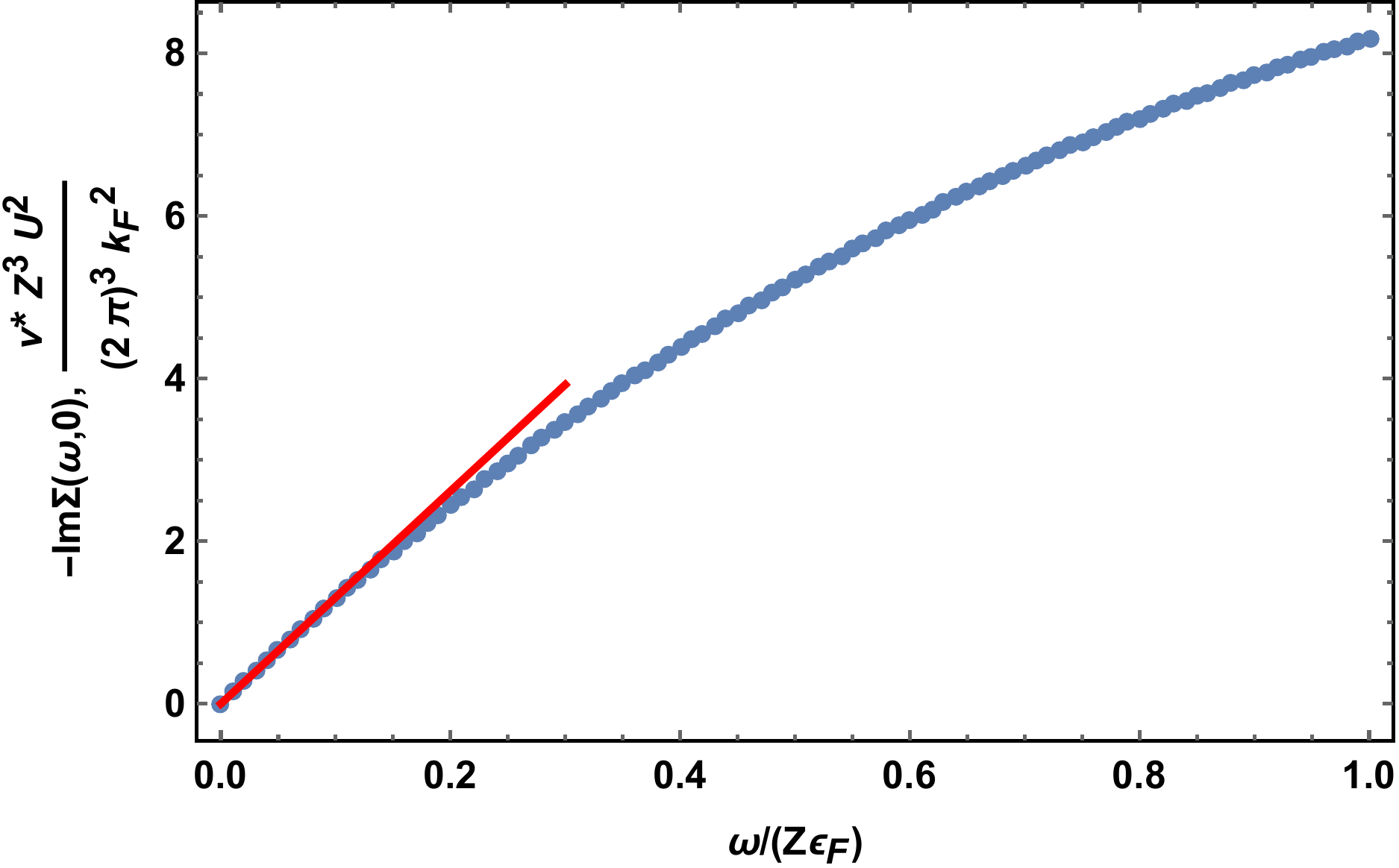} \quad
\includegraphics[width=0.464\linewidth]{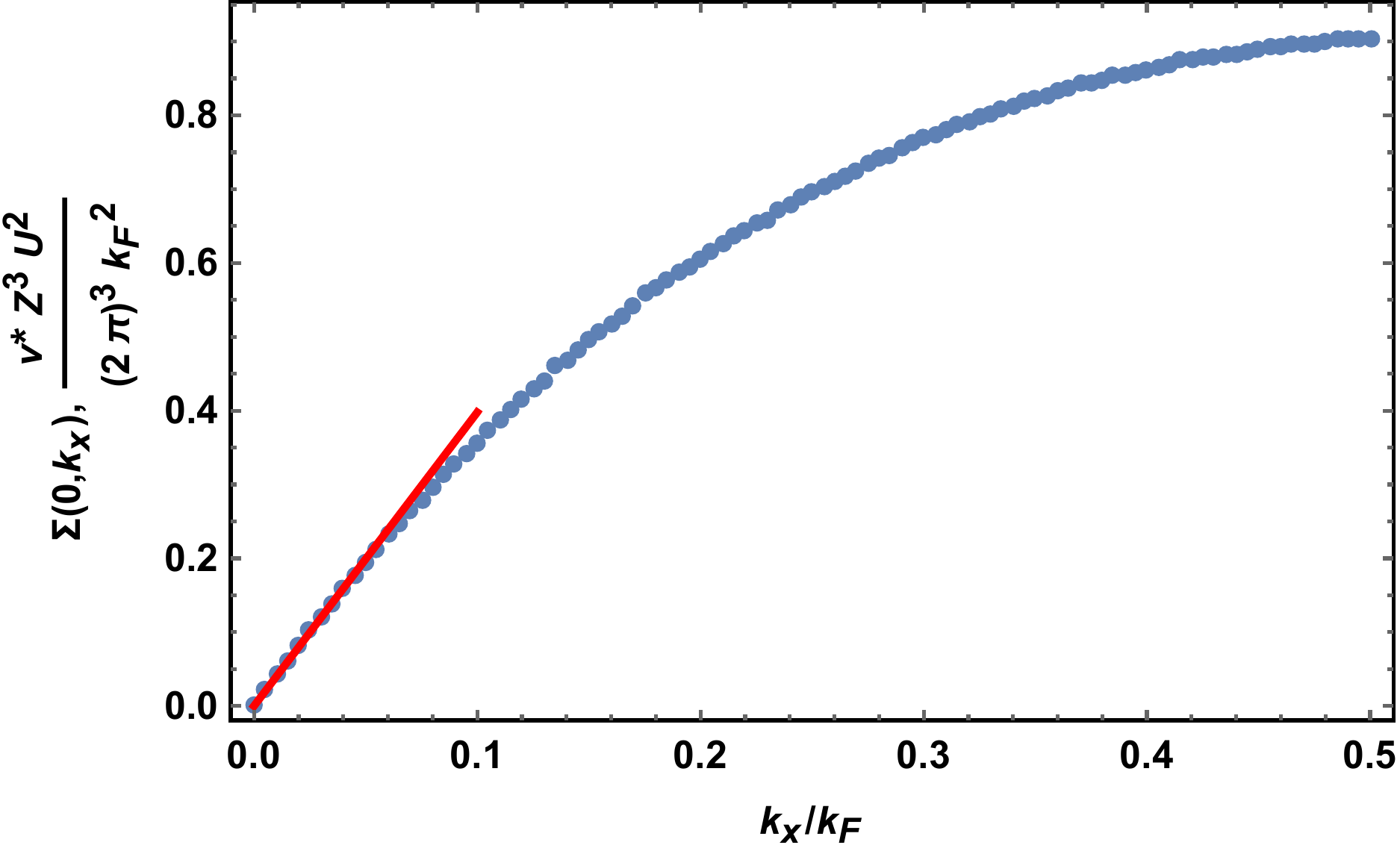}
\caption{\small \label{fig:Sigma_FL}  {The self-energy (\ref{Sigma_IR}) in the Fermi liquid regime for a circular Fermi surface. Left panel: frequency dependence of the self-energy. Right panel: momentum dependence of the self-energy. The red lines shows the fit with $\alpha_1 = 13.1$ for frequency and $\alpha_2 \approx 4$ for the momentum dependencies of the self-energy. The momentum dependence of the self-energy is shifted by a constant in Eq.~\eqref{app:eq_sigma_kx}, so that it originates from $k_x=0$.}}
\end{figure}

\subsection{Non-Fermi liquid at $\varepsilon_F^2/U\ll T\ll U$}\label{app:high_T}

At higher temperatures than the renormalized Fermi energy, one expects that the kinetic energy of the fermions becomes negligible. In this Subsection, we show that the system is a local NFL.

We assume and then verify that in area $\sim k_F^2$ around the Fermi surface, at higher-temperatures the Green's function to zeroth order takes a $\bf k$-independent form
\be
G(\omega, {\bf k}) \simeq G^{(0)}(\omega) + G^{(1)}(\omega, {\bf k}) + \cdots,~~~G^{(1)}\ll G^{(0)}.
\ee
The derivation of $G^{(0)}$ is quite similar to that in the 0d SYK model, the only difference being an momentum integral that cancels the $k_F^{-2}$ in the coupling constant in our case. Therefore, at leading order, the fermion Green's function is
\begin{align}
    G^{(0)}(\omega) \simeq -i \pi^{1/4} \frac{\mathrm{sgn}(\omega)}{\sqrt{U|\omega|}}
\end{align} 
with the fermionic self-energy given by
\be
\Sigma^{(0)}(\omega)\simeq -i\pi^{-1/4}\sqrt{U|\omega|}\sgn(\omega).
\ee
This solution is valid when $\omega,\varepsilon_F\ll \Sigma(\omega)$, which is satisfied in the temperature range
\begin{align}
 \frac{\varepsilon_F^2}{U}\ll T \ll U,   
\end{align}
which is complementary to the temperature regime for heavy Fermi liquid behavior.

The momentum dependence of the self-energy in this regime can be evaluated perturbatively: 
\begin{align}
    \Sigma(x) \simeq & \Sigma^{(0)}(x) + \Sigma^{(1)}(x), \\
   k_F^4 \Sigma^{(0)}(x) = & - U^2 G^{(0)}(x)^2 G^{(0)}(-x) \\ 
  k_F^4 \Sigma^{(1)}(x) = & -\mathcal{J}^2 G^{(0)}(x)^2 G^{(1)}(-x) - 2 \mathcal{J}^2 G^{(0)}(x) G^{(0)}(-x) G^{(1)}(x) - 3\Delta_{4e}^2 G^{(0)}(-x)^2 G^{(1)}(-x). \label{app:G_1}
\end{align}
The corresponding Green's function is \cite{Chowdhury2018Translationally}
\begin{align}
    G(\omega,\mathbf{k}) \simeq& G^{(0)}(\omega) + G^{(1)}(\omega, \mathbf{k}),\\ 
    G^{(0)}(\omega) =& -\frac{1}{\Sigma^{(0)}(\omega)},\\ 
    G^{(1)}(\omega,\mathbf{k}) =& G^{(0)}(\omega)^2 \( \xi_{\bf k} + \Sigma^{(1)}(\omega, {\bf k}) \). 
\end{align}
{The perturbation series is controlled by the small parameter $\varepsilon_F/U$.}

{Applying the Fourier transform to the self-energy correction \eqref{app:G_1}, we notice that only $G^{(1)}$ among the product of three Green's functions explicitly depends on momenta. As such,
\beq 
\Sigma^{(1)}(\omega,\mathbf{k}) \sim \int d\mathbf{q} \, d\Omega \, f(\Omega) G^{(1)}(\Omega-\omega,\mathbf{k}-\mathbf{q}),
\eeq
where $f(\Omega)$ is defined by $G^{(0)}(\Omega)$ which is momenta independent.}
The key observation here is, since the external momentum $\bf k$ can be absorbed into the internal momentum $\bf Q$, the mometum integral is independent on $\bf k$. Therefore, we have
\be
\Sigma^{(1)}(\omega,\mathbf{k}) =  \Sigma^{(1)}(\omega),
\ee
and the $\bf k$ dependence of $\Sigma(\omega, \mathbf{k})$ comes in the perturbative expansion at the next order. 
Thus by dimensional analysis,
\be
\Sigma(0,{\bf k}) \sim \frac{\varepsilon_F}{U}{\bf v}_F\cdot {\bf k}.
\ee

\section{Ward identity and superfluid density}\label{app:ward}

In this Appendix we derive the modified Ward identity for superconductors.

\subsection{Ward identity in a metal}\label{app:ward_metal}

For completeness, we first derive the Ward idetity from the path integral formalism for a metal with free-electron dispersion. 
This derivation is well-known and presented in detail in quantum field theory books, e.g.,~\cite{Peskin2018Introduction}.

The Ward identity, introduced in the main text in Eq. (\ref{eq:34}), directly relates the quasiparticle current vertex $\mathbf{\Gamma}$ to the quasiparticle velocity $-\partial_\mathbf{k} G^{-1}$:
\be
\mathbf{\Gamma}(k,k) = -\partial_\mathbf{k} G^{-1}(k).
\ee
This is a direct consequence of minimal coupling to the electromagnetic field, which in turn comes from gauge invariance as we show below.

We begin with the path integral for two-point functions
\begin{align}
{\langle \psi(x_1) \bar{\psi}(x_2) \rangle} \sim & \int \mathcal{D}\psi \mathcal{D}\bar{\psi} \, \psi(x_1) \bar{\psi}(x_2) e^{-S[\psi, \bar{\psi}]} \nonumber \\=& \int \mathcal{D}\psi' \mathcal{D}\bar{\psi}' \, \psi'(x_1) \bar{\psi}'(x_2) e^{-S[\psi', \bar{\psi}']}, \label{W12}
\end{align}
where in the second line we have performed a change of variable  $\psi \to \psi'= e^{i\alpha(x)} \psi$, where $\alpha(x)$ is an arbitrary function of $x=(\tau, \mathbf{r})$.

Expanding the bottom line in Eq.~(\ref{W12}) to linear order in $\alpha(x)$ and subtracting it from the top line in the same equation we get
\be
0= \int \mathcal{D}\psi \mathcal{D}\bar{\psi} \int_x \, i\alpha(x)\psi(x_1) \bar{\psi}(x_2) \[\delta(x-x_1)-\delta(x-x_2) -i\partial_\mu j^\mu(x) \] e^{-S[\psi, \bar{\psi}]},
\label{eq:2}
\ee
where $\partial_\mu = (\partial_\tau, \partial_\mathbf{r})$ and $j^\mu = (i\bar{\psi}\psi, -i(\bar{\psi}\partial_\mathbf{r}\psi - (\partial_\mathbf{r}\bar{\psi})\psi )/(2m))$.
Here we have used the conservation law due to $U(1)$ symmetry of the action. Namely, the change in the action due to $\psi \to \psi'$ can only be proportional to the gradient term $\partial_\mu \alpha(x)$, with the coefficient being the conserved current as stated by the Noether's theorem.

We then obtain the following relation between correlation functions
\be
-i \partial_\mu \langle j^\mu (x) \psi(x_1) \bar{\psi}(x_2) \rangle =- [\delta(x-x_1)-\delta(x-x_2)] \langle \psi(x_1) \bar{\psi}(x_2) \rangle. \label{app:jpp}
\ee
{We apply} the Fourier operator
$\mathcal{\hat F}\circ = \int_{x, x_1, x_2}  e^{{+}ipx} e^{{+}ikx_1} e^{{-}i(k+p)x_2} \circ$ to both sides of Eq.~(\ref{app:jpp}) { and take the limit of zero frequency $\omega \to 0$ in the Fourier components of the particle current $j^\mu(\omega, \mathbf{p})$}. {Then using the Wick's theorem 
for the left side of Eq.~(\ref{app:jpp})}, we obtain the equation (\ref{eq:ward}) from the main text
\be
{\mathbf{p} \cdot \mathbf{\Gamma}(k,k+p)} G(k) G(k+p) = {-}G(k) {+} G(k+p), \label{app:ward_id}
\ee
where we introduced the current vertex function $ \mathbf{\Gamma}(k,k+p) = (\mathbf{k} + \mathbf{p}/2)/m$ for a free fermion system. 
{The Green's function is defined as $G(k) = -\langle \psi_k \bar{\psi}_k \rangle$.}
Taking the limit $\mathbf{p}\to 0$ in Eq.~(\ref{app:ward_id}), we derive the current vertex to quasiparticle velocity relation (\ref{eq:34}) used in the main text.

\subsection{Modified Ward identity and superfluid density in a charge-$2e$ superconductor}\label{app:ward_BCS}

For a superconductor (charge-$2e$ or charge-$4e$), $U(1)$ symmetry is broken, and Ward identity can be modified accordingly. For completeness, here we derive the modified Ward identity in a mean-field BCS superconductor.

In a BCS superconductor, the action contains the anomalous term
\be
S \supset { -} \int_x \Delta \bar\psi_{\uparrow} (x)  \bar\psi_{\downarrow} (x) + h.c.,
\ee
which does not remain invariant under $\psi_{\uparrow,\downarrow}\to \psi_{\uparrow,\downarrow}'= e^{i\alpha(x)} \psi_{\uparrow,\downarrow}$. 

To account for the change of the action due to the anomalous term, Eq.~\eqref{eq:2} should be modified to
\begin{align}
0=& \int \mathcal{D}\psi \mathcal{D}\bar{\psi} \int_x \, i\alpha(x)\psi_{\uparrow}(x_1) \bar{\psi}_{\uparrow}(x_2) \[\delta(x-x_1)-\delta(x-x_2) -i\partial_\mu j^\mu(x) \right.\nonumber\\
&~~~~~~~~~~~~~~~~~~~~~~~~~~~\left.-2\Delta \bar\psi_{\uparrow}(x) \bar\psi_{\downarrow}(x) +2\Delta  \psi_{\uparrow}(x) \bar\psi_{\downarrow}(x) \] e^{-S[\psi, \bar{\psi}]},
\label{app:S_var_BCS}
\end{align}
Similar to the previous section, the equation (\ref{app:S_var_BCS}) translates to momentum representation as 
\begin{align} \nonumber
   & {\mathbf{p} \cdot \mathbf{\Gamma}(k,k+p)} {G}(k)  {G}(k+p)  
   ={-}  {G}(k)+  {G}(k+p) \nonumber\\
    &+  \hat{\mathcal{F}} \circ 2  \Delta \langle \bar\psi_{\uparrow}(x) \bar\psi_{\downarrow}(x) \psi_{\uparrow}(x_1) \psi_{\uparrow}^\dag(x_2) \rangle - \hat{\mathcal{F}} \circ 2  \Delta \langle \psi_{\downarrow}(x) \psi_{\uparrow}(x) \psi_{\uparrow}(x_1) \psi_{\uparrow}^\dag(x_2) \rangle ,
\end{align}

Applying the Wick's theorem {and assuming the $s$-wave real constant pairing}, we have
\begin{align}
\hat{\mathcal{F}} \circ 2  \Delta \langle \bar\psi_{\uparrow}(x) \bar\psi_{\downarrow}(x) \psi_{\uparrow}(x_1) \psi_{\uparrow}^\dag(x_2) \rangle = 2 G(k) \Sigma_{\Delta}(k)G(k+p),
\label{app:ward_BCS_trace}
 \end{align}
 whee $\Sigma_{\Delta}(k)$ is the self-energy due to the pairing vertex.
For a BCS superconductor, we have
\be
 G (k) = - \frac{i\omega + \xi_{\bf k}}{\omega^2+\xi^2_{\bf k}+\Delta^2},~~~\Sigma_\Delta(k) = \frac{\Delta^2}{i\omega+\xi_{\bf k}}.
 \label{app:Gorkov}
\ee
where $\xi_{\bf k}=({\bf k}-e{\bf A})^2/2m-\mu$ is the dispersion relation. Taking the $\mathbf{p} \to 0$ limit, we have the modified Ward identity (\ref{eq:ward}) for a superconductor
\begin{align} 
 &
 \mathbf{\Gamma}(k,k) {G}^2(k)=   \partial_\mathbf{k} G({k}) -2 G^2(k)\partial_{\bf k} \Sigma_{\Delta}(k). \label{app:ward1}
\end{align}
 
The first term in Eq.~\eqref{app:ward1} again cancels the diamagnetic contribution {discussed in Section \ref{sec:S_density} in the main text}, and the second term yields
\be
n_s = 4\int_k \frac{\mathbf{k}^2}{m}\frac{\Delta_0^2}{(\omega^2+\xi_\mathbf{k}^2+\Delta^2)^2}, \label{app:n_BCS}
\ee
where the additional factor of 2 accounts for both spin species.
{Introducing the density of states for a two-dimensional case $\nu_0 = k_F^2/(2\pi \varepsilon_F)$ at zero temperature, we have 
\beq
n_s=4 \int_k \frac{\mathbf{k}^2}{m}\frac{\Delta_0^2}{(\omega^2+\xi_\mathbf{k}^2+\Delta^2)^2} = \frac{4 \nu_0 \varepsilon_F}{\pi} \int_{-\infty}^{+\infty}\!\! d\omega \int_{-\infty}^{+\infty}\!\! d\xi \, \frac{\Delta_0^2}{(\omega^2+\xi^2+\Delta^2)^2} = 2 \times \frac{k_F^2}{\pi} = n.
\eeq
}
 
The fact that $n_s$ is independent of $\Delta$ is related to the superfluid density being a Fermi surface effect. No matter how small $\Delta$ is, pairing always equally  mixes electrons and holes at the Fermi surface. Therefore $n_s$ is significant even if $\Delta$ is small. 

\subsection{Modified Ward identity in a charge-$4e$ superconductor}\label{app:ward_4e}

Now, we shall derive the Ward identity for a charge-$4e$ superconductor from the main text \eqref{eq:20}. The imaginary time action for a charge-$4e$ superconductor reads 
\begin{align}
    S \supset &  \int_x \Big( \sum_{i=1}^N \Psi^\dag_{x i} \sigma_0 \left(\partial_\tau-\frac{\partial_\mathbf{r}^2}{2m}-\mu \right) \Psi_{x i} + \frac{k_F^{-d}}{N^{3/2}} \sum_{i<j,k<l,i<k}^N \left( {\Delta_{4e}}_{,ij;kl} \, \Psi^\dag_{x i} i \sigma_y (\Psi_{x j}^\dag)^T  \,  \Psi^\dag_{x k} i \sigma_y (\Psi_{x l}^\dag)^T  + h.c.\right) \Big), \label{app:S_charge4e}
\end{align}
where we use the fermionic spinor $\Psi_{x i}^\dag = \begin{pmatrix} \bar{\psi}_{i\uparrow}(x) & \bar{\psi}_{i\downarrow}(x) \end{pmatrix}$. 

Similar to Appendix \ref{app:ward_metal}, we define the two-point function for a given spin and flavour
\begin{align}
  \langle \psi_{1 \uparrow}(x_1) \bar{\psi}_{1 \uparrow}(x_2) \rangle \sim \int \mathcal{D} \psi \mathcal{D} \bar{\psi} \, e^{-S[\psi,\bar{\psi}]} \psi_{1 \uparrow}(x_1) \bar{\psi}_{1 \uparrow}(x_2),
\end{align}
which remains invariant under the change of variables $\psi_{n s} \to \psi_{n s}' = e^{i \alpha(x)} \psi_{n s}$.
Accordingly, the Ward identity is modified as 
\begin{align} \nonumber
   &-i \partial_\mu \langle j^\mu (x) \psi_{1 \uparrow}(x_1) \bar{\psi}_{1 \uparrow}(x_2) \rangle =- [\delta(x-x_1)-\delta(x-x_2)] \langle \psi_{1 \uparrow}(x_1) \bar{\psi}_{1 \uparrow}(x_2) \rangle \\ \nonumber & -  \frac{4  k_F^{-d}}{N^{3/2}} \,  \sum_{i<j,k<l,i<k}^N {\Delta_{4e}}_{,ij;kl} \,  \langle \psi_{1 \uparrow}(x_1) \bar{\psi}_{1 \uparrow}(x_2)  \Psi^\dag_{x i} i \sigma_y (\Psi_{x j}^\dag)^T  \,  \Psi^\dag_{x k} i \sigma_y (\Psi_{x l}^\dag)^T \rangle \\ &+ \frac{4 k_F^{-d}}{N^{3/2}} \,  \sum_{i<j,k<l,i<k}^N {\Delta_{4e}}_{,ij;kl} \,  \langle \psi_{1 \uparrow}(x_1) \bar{\psi}_{1 \uparrow}(x_2) \Psi^T_{x l} i \sigma_y^T \Psi_{x k}  \,  \Psi^T_{x j} i \sigma_y^T \Psi_{x i} \rangle, \label{app:ward_id_4e_x}
\end{align}
where the current is 
\be
j^\mu(x) = \sum_{i=1}^N\sum_{\sigma=\uparrow,\downarrow}(i\bar{\psi}_{i\sigma}\psi_{i\sigma}, -\frac{i}{2m}(\bar{\psi}_{i\sigma}\partial_\mathbf{r}\psi_{i\sigma} - (\partial_\mathbf{r}\bar{\psi}_{i\sigma})\psi_{i\sigma} )).
\ee

Let's evaluate the $6$-fermion correlation function in Eq.~\eqref{app:ward_id_4e_x} using that under disorder average $\langle {\Delta_{4e}}_{,ij;kl} {\Delta_{4e}}_{,i'j';k'l'} \rangle = \Delta_{4e}^2 \delta_{ii'}\delta_{jj'}\delta_{kk'}\delta_{ll'}$:
\begin{align} \nonumber
    &-  \frac{4  k_F^{-d}}{N^{3/2}} \,  \sum_{i<j,k<l,i<k}^N {\Delta_{4e}}_{,ij;kl} \,  \langle \psi_{1 \uparrow}(x_1) \bar{\psi}_{1 \uparrow}(x_2)  \Psi^\dag_{x i} i \sigma_y (\Psi_{x j}^\dag)^T  \,  \Psi^\dag_{x k} i \sigma_y (\Psi_{x l}^\dag)^T \rangle \\  &=  \frac{4  k_F^{-2d}}{N^3} \Delta_{4e}^2 \,  \sum_{i<j,k<l,i<k}^N  \langle \psi_{1 \uparrow}(x_1) \bar{\psi}_{1 \uparrow}(x_2)  \Psi^\dag_{x i} i \sigma_y (\Psi_{x j}^\dag)^T  \,  \Psi^\dag_{x k} i \sigma_y (\Psi_{x l}^\dag)^T 
    \int_{x'} \Psi^T_{x' l} i \sigma_y^T \Psi_{x' k}  \,  \Psi^T_{x' j} i \sigma_y^T \Psi_{x' i} \rangle. \label{app:6f}
\end{align}
Applying the Wick's theorem to the equation above, we notice that in the $i$th spinor the flavour is fixed to $i=1$ and only the $\uparrow$-component of it contributes to the Ward identity. Therefore, the $6$-fermion function \eqref{app:6f} becomes
\begin{align} \nonumber
    &\frac{4  k_F^{-2d}}{N^3} \Delta_{4e}^2 \,  \sum_{j,k,l=1}^N \int_{x'} \langle \psi_{1 \uparrow}(x_1) \bar{\psi}_{1 \uparrow}(x_2) \psi_{1 \uparrow}(x') \bar{\psi}_{1 \uparrow}(x)  \rangle \langle \psi_{j \downarrow}(x') \bar{\psi}_{j \downarrow}(x) \rangle  \langle \psi_{k \downarrow}(x') \bar{\psi}_{k \downarrow}(x) \rangle \langle \psi_{l \uparrow}(x') \bar{\psi}_{l \uparrow}(x) \rangle \\ &= 4 \Delta_{4e}^2 \,  \int_{x'} {\Sigma_\Delta}_{\uparrow\uparrow}(x,x') \langle \psi_{1 \uparrow}(x_1) \bar{\psi}_{1 \uparrow}(x_2) \psi_{1 \uparrow}(x') \bar{\psi}_{1 \uparrow}(x)  \rangle.
\end{align}

Thus, the Ward identity for our charge-$4e$ superconductor reads  
\begin{align} \nonumber
   &-i \partial_\mu \langle j^\mu (x) \psi_{1 \uparrow}(x_1) \bar{\psi}_{1 \uparrow}(x_2) \rangle =- [\delta(x-x_1)-\delta(x-x_2)] \langle \psi_{1 \uparrow}(x_1) \bar{\psi}_{1 \uparrow}(x_2) \rangle \\ \nonumber 
   &+ 4 \Delta_{4e}^2 \,  \int_{x'} {\Sigma_\Delta}_{\uparrow\uparrow}(x,x') \langle \psi_{1 \uparrow}(x_1) \bar{\psi}_{1 \uparrow}(x_2) \psi_{1 \uparrow}(x') \bar{\psi}_{1 \uparrow}(x)  \rangle \\
   &- 4 \Delta_{4e}^2 \,  \int_{x'} {\Sigma_\Delta}_{\uparrow\uparrow}(x'\!,x) \langle \psi_{1 \uparrow}(x_1) \bar{\psi}_{1 \uparrow}(x_2) \psi_{1 \uparrow}(x) \bar{\psi}_{1 \uparrow}(x')  \rangle. \label{app:ward_id_4e_x_S}
\end{align}

We apply the Fourier transform $\mathcal{\hat F}\circ = \int_{x, x_1, x_2}  e^{+ipx} e^{+ikx_1} e^{-i(k+p)x_2} \circ$ to the both sides of Eq.~\eqref{app:ward_id_4e_x_S} and take $\omega \to 0$ limit:
\begin{align} \nonumber
   &   \int_q \frac{{\bf p} \cdot\({\bf q}+{\bf p}/2\)}{m}\langle \psi_{1 \uparrow}(k) \bar{\psi}_{1 \uparrow}(q) \psi_{1 \uparrow}(q+p) \bar{\psi}_{1 \uparrow}(k+p)\rangle = - \langle \psi_{1 \uparrow}(k+p) \bar{\psi}_{1 \uparrow}(k+p) \rangle  +  \langle \psi_{1 \uparrow}(k) \bar{\psi}_{1 \uparrow}(k) \rangle \\ \nonumber 
   &- 4\Delta_{4e}^2  \int_q {\Sigma_\Delta}_{\uparrow\uparrow}(q+p) \langle \psi_{1 \uparrow}(k) \bar{\psi}_{1 \uparrow}(q) \psi_{1 \uparrow}(q+p) \bar{\psi}_{1 \uparrow}(k+p) \rangle \\ 
   &+ 4\Delta_{4e}^2  \int_q {\Sigma_\Delta}_{\uparrow\uparrow}(q) \langle \psi_{1 \uparrow}(k) \bar{\psi}_{1 \uparrow}(q) \psi_{1 \uparrow}(q+p) \bar{\psi}_{1 \uparrow}(k+p)\rangle.
\end{align}
Using the diagrammatic technique in Matsubara time with $G(k) = - \langle \psi_{1 \uparrow}(k) \bar{\psi}_{1 \uparrow}(k) \rangle$ and $\Sigma_{\Delta_{4e}} \equiv {\Sigma_\Delta}_{\uparrow\uparrow}$, we obtain Eq.~\eqref{eq:21} and Fig.~\ref{fig:diagrams} in the main text.

\end{appendix}

\end{widetext}

\bibliography{refs}

\end{document}